\documentclass{article}

\usepackage{makecell}
\usepackage{chemfig}
\usepackage{chemformula}
\usepackage{caption,subcaption}
\usepackage{booktabs}
\usepackage{booktabs, tabularx}
\newcolumntype{C}{>{\centering\arraybackslash}X}
\usepackage{adjustbox}
\usepackage{adjustbox}
\usepackage[version=3]{mhchem}
\usepackage[utf8x]{inputenc}

\usepackage{mathtools}
\usepackage{amsmath, amssymb, graphics, setspace}
\usepackage{comment}
\usepackage{tabularx}

\usepackage[T1]{fontenc}    
\usepackage{hyperref}       
\usepackage{url}            
\usepackage{booktabs}       
\usepackage{amsfonts}       
\usepackage{nicefrac}       
\usepackage{microtype}      
\usepackage{graphicx}
\usepackage{doi}
\usepackage[numbers]{natbib}
\usepackage{xcolor}
\hypersetup{
    colorlinks,
    linkcolor={red!50!black},
    citecolor={blue!50!black},
    urlcolor={blue!80!black}
}

\title{Study of Interfacial Rheology of Human Serum Albumin Microcapsules using Electrodeformation Technique}
\author{Sneha Puri and 
	Rochish M. Thaokar \\
	 E-mail: rochish@che.iitb.ac.in\\
 Phone: +91 (22) 2576 7241
Fax: +91 (22) 2572 6895\\
	Department of Chemical Engineering,\\ Indian  Institute of  Technology Bombay, Mumbai-400076 }

\begin{document}
\maketitle

\begin{abstract}
Understanding the mechanical characterization of microcapsules is critical for their precise applications, such as in pharmaceuticals, cosmetics, agriculture, food industries, etc. Microcapsules synthesized from different materials show distinct mechanical characteristics.  It is therefore necessary  to study these systems considering their respective micro-physics for an exact theoretical model suitable to these systems. In the present work, we conducted mechanical characterization of the membrane of human serum albumin (HSA) microcapsules using the electrodeformation technique. Proteins are widely used as encapsulating materials for many biomedical and food industries. Relating the microstructure to the mechanical properties of a protein-membrane is a challenge owing to its complex structure and sensitivity towards different environmental conditions such as pH, temperature, and solvents. Interfacial rheology of human serum albumin microcapsules studied using the electrodeformation technique shows that HSA capsules are strain-softening in nature. The viscoelasto-electrohydrodynamic model was utilized to understand the creep mechanism in the human serum albumin capsules. The effect of different reaction parameters such as protein concentration and pH on the morphology of the capsule membrane was investigated, and an attempt has been made to correlate microstructure with the mechanical properties. The pH has a remarkable effect on the morphology of HSA microcapsules which is also reflected in their mechanical characteristics. The capsule synthesized with carbonate buffer shows very distinct morphology with pores on the membrane surface, making the membrane less elastic with significant nonrecoverable creep. The capsules synthesized with different protein concentrations at the same pH condition show different morphology and thus different rheological properties. Capsules with a low concentration of HSA show smooth membrane structure with higher Young's modulus than the capsules synthesized at a very high concentration which show a rough folded wavy structure with low membrane elastic modulus. The effect of frequency on the interfacial rheological properties of human serum albumin capsules was studied using a frequency sweep test using the electrodeformation technique. The rheological properties were computed incorporating the viscoelasto-electrohydrodynamic model for the oscillatory response of capsules. The results show that the elastic response dominates in the high-frequency regime. Thus the electrodeformation technique allows studying the effect of very high-frequency 1 Hz to 1 kHz, which is otherwise not possible with the conventional rheometers.

\end{abstract}


\section{Introduction}

A microcapsule is a liquid droplet covered with a thin sheet (shell) of self-assembled or crosslinked particles or polymers of a few 10s of microns in size. Common examples of natural microcapsules include red blood cells, fat globules in milk, etc. Synthetic microcapsules made up of crosslinked polymers find tremendous use in several applications. Specifically, crosslinked proteins as the shell material have attractive applications in the biomedical, pharmaceuticals, cosmetics and food industries. This paper demonstrates an electrodeformation
technique to study the interfacial rheology of human serum albumin microcapsules. HSA
is a biocompatible protein and having a very attractive binding capacity to many drugs.
Below we briefly discuss the studies on the characterization of human serum albumin microcapsules.

The fourier transform infrared (FTIR) spectroscopic studies on HSA capsules synthesized by the emulsification method showed the participation of amino, hydroxyl, and carboxyl functional groups of albumin in the formation of the crosslinked HSA microcapsules \citep{levy1991fourier}. The extent of involvement of these functional groups depends on the reaction conditions during the capsule synthesis process. \citet{levy1991fourier} studied the chemical characterization of HSA capsules using  FTIR spectroscopy and TNBS titration techniques. Their method was focused on the determination of the amount of free amino groups of the microcapsules, using the TNBS titration method to understand the degree of crosslinking, and the FTIR technique for the involvement of various functional groups at different reaction conditions.

Proteins are known to be sensitive to pH and acquire a configuration that depends upon the protonation/deprotonation of amino groups and ionization of their carboxylic group present on the amino acid side groups \citep{dumetz2008effects}. At low pH, the involvement of only free amine groups is seen, whereas, for high pH, carboxyl and hydroxyl groups also participate in the reaction. Moreover, at a higher pH, the ionization of the amine groups is suppressed, thereby favoring the amide reaction with the reducing agent. HSA capsules synthesized with phosphate buffer in the pH range between 5.9 to 8 yielded larger-sized microcapsule aggregates with smooth morphology that collapsed after drying. However, with a similar reaction condition, microcapsules prepared with a carbonate buffer (pH 9 to 11) resulted in smaller, pale yellow, rough individual spherical capsules \citep{levy1991fourier}. 

In yet another study, it was found that the acylation of amino groups of HSA is more intense with acetate buffer than with the phosphate buffer at the same pH of 5.9 \citep{edwards1993determination}. This gives the advantage of increasing crosslinking density of the membrane of HSA microcapsules without raising the pH. The difference between acetate and carbonate buffers was attributed to the difference in the ionic concentration of the two buffers. The exact reason for why the higher ionic concentration of the acetate buffer favors acylation of the amino groups at the same pH though is unclear.

Increasing reaction time shows a similar effect as upon increase in pH on membrane morphology of the capsules. At pH 9.8 (carbonate buffer) with 2 min reaction time, bigger $\sim$ 30 $\mu m$ capsules with smooth membrane morphology is formed, while as the reaction time increases, smaller microcapsules with rough morphology are obtained. Initially, only free amines participate in the reaction, and as the reaction time increases, more functional groups are involved in the reaction producing rough and more granular microcapsules \citep{levy1995fourier}. Similarly, with the increase in the concentration of crosslinker, a rough membrane is formed, showing a higher degree of crosslinking \citep{levy1995fourier,levy1991fourier}.  

 To summarize, the literature survey about the synthesis of HSA capsules indicates that two types of capsules can be synthesized by tuning the different reaction parameters.
 \begin{enumerate}
 
 \item Poorly crosslinked type-1 capsules that show smooth membrane morphology and involve mainly the reaction of free amines; can be synthesized either at low pH, at low TC concentration (2.5 percent), or with a short reaction time at high pH. The extent of crosslinking is much lower in this case.
 \item Highly crosslinked type-2 capsules are smaller in size with rough morphology and involve the formation of more $\beta$ sheets, ester, and anhydride bonds and can be synthesized by intensifying the reaction by increasing one of the reaction parameters; pH, reaction time, or TC concentration.
  \end{enumerate}

Designing a microcapsule with desired mechanical characteristics is a challenge in microencapsulation technology. The capsules prepared with different materials, such as chitosan, alginate, and proteins, show distinct membrane mechanical properties. For example, HSA-alginate capsules show strain hardening \citep{carin2003compression} behavior while serum albumin (HSA/BSA) capsules exhibit strain-softening nature \citep{gires2014mechanical, gubspun2016characterization}. The reaction parameters and fabrication techniques can also significantly affect the membrane properties of microcapsules synthesized with the same system. Therefore it is often necessary to study their properties on a case-by-case basis to understand their characteristics that are relevant to their applications.

Mechanical characterization of water-in-water albumin capsules using the microfluidic technique has been reported in several studies \citep{de2014mechanical,de2015stretching,gires2016transient, gubspun2016characterization}. Various theoretical models such as Hooke's law, Neo-Hookean, and Skalak law describe the 2D membrane characteristics of elastic microcapsules. Most of the literature studies considered strain-softening Neo-Hookean law to compute mechanical properties of albumin capsules \citep{gubspun2016characterization,gires2014mechanical}, barring the study about stretching of HSA capsules in planar elongation flow by \citet{de2015stretching} which reports that these capsules follow generalized Hooke's law with Poisson's ratio of 0.4  \citep{de2015stretching}. 

 The deformation study of HSA capsules in elongational flow and the atomic force microscopy (AFM) technique allowed the estimation of surface shear modulus and 3-D Young's modulus in the small deformation limit. A combination of these two methods revealed that Young's modulus and membrane thickness both increase with the HSA concentration and size of the capsule \citep{de2014mechanical}. Studies on the influence of protein concentration and the size of the microcapsule on the mechanical properties of the HSA microcapsule membrane using the microfluidic technique suggest that as the size and the HSA concentration increase, the shear modulus also increases \citep{gires2014mechanical,gubspun2016characterization,de2016tank}. Simultaneous fabrication and characterization of HSA microcapsules using a microfluidic (double flow-focusing setup) technique have been reported in the literature; however, it is difficult to remove the excess unreacted precursor and to make stable capsules with this technique. The study also showed that as the reticulation time increases, the shear modulus also increases \citep{chu2013fabrication}.

  Studies have been conducted on ovalbumin capsules to understand their mechanical characteristics. The study of the deformation of ovalbumin capsules under shear flow through a cylindrical microchannel at different flow rates showed that the use of a Neo Hookean law gave a constant value of shear modulus in the low deformation regime \citep{lefebvre2008flow}. The study comparing the mechanical and chemical characterization of ovalbumin capsules, done using the microfluidic technique and combined with determining free amino groups using the TNBS method, shows that the shear modulus and the amino groups are nearly constant with the reaction pH for the capsules fabricated after 5 min of reticulation. The shear modulus increases with the reaction time, while the $NH_{2}$ content decreases with it, indicating the conversion of progressive amines to amides. An overall increase in shear modulus with pH is also observed for ovalbumin capsules, similar to the HSA capsules \citep{chu2011comparison}, but with an unexpected rise in $NH_{2}$ content. 
  
 Very few studies have been conducted on the viscoelastic characterization of microcapsules. The theoretical studies considered the Kelvin-Voigt model to describe the viscoelastic properties of the membrane under shear flow \citep{barthes1985role,yazdani2013influence}. An experimental study was performed by \citet{chien1978theoretical} on the erythrocyte membrane in which the Kelvin-Voigt model was used to find the membrane viscosity using the micropipette technique. Recently, alginate-coated chitosan capsules were studied using the microelectromechanical (MEMS) microgripper technique, and the Kelvin model was used to find the viscoelastic characteristics of these microcapsules \citep{kim2009elastic}. The viscoelastic properties of oil-in-water vitamin A capsules coated with starch were studied using the four-element Burger model \citep{zhang2019effect}. Similarly, urea-formaldehyde microcapsules showed viscoelastic characteristics, and the relaxation was described by the three-element Maxwell  model \citep{han2019investigation}. Recently, a novel method has been developed, by integrating a deep convolutional neural network with a high-fidelity mechanistic capsule model, to predict the membrane viscosity and elasticity of a microcapsule from its dynamic deformation when flowing in a branched microchannel \citep{lin2021high}.

Only a few studies have been reported on the viscoelastic properties of water-in-water HSA microcapsules.
  The study of tank treading motion of  HSA microcapsules indicates that these capsule membranes are viscoelastic in nature, and the membrane viscosity was computed from the measurement of the period of rotation of the membrane \citep{de2016tank}. The transient response of cross-linked HSA capsules was studied by flowing capsules through a sudden expansion, and comparing the characteristic time scales of the capsule relaxation with a complete numerical model of the relaxation of a capsule flowing in a rectangular channel. The study shows that the crosslinked HSA membrane is viscoelastic and that the relaxation is solely a function of the ratio of the relaxation time to the flow convective time \citep{gires2016transient}.

In the present work, we use the electrodeformation technique to study the viscoelastic properties of HSA microcapsules using creep and oscillatory tests. An elasto-electrohydrodynamic model is used to compute the membrane elasticity and viscosity by corelating the fitted parameters of five element spring dashpot model. The dynamics of the deformation of the capsule was studied at constant electric stress for different frequencies to study the effect of frequency on membrane rheological properties. The electrodeformation method allows interfacial rheological studies at very high frequencies, not easily accessible in conventional interfacial rheometers.

 \section{Materials and methods}
 Human serum albumin protein standard (P8119), span 85, chloroform, terepthyalyolchloride, and silicone oil (350 cSt) were obtained from Sigma Aldrich. Cyclohexane was obtained from Imparta chemicals, and pH 7.4 PBS buffer from LOBA chemicals. All the chemicals were used as procured.  
 
 \begin{figure}
	\centering
 	\begin{subfigure}{0.7\linewidth}
 	\includegraphics[width=1\linewidth]{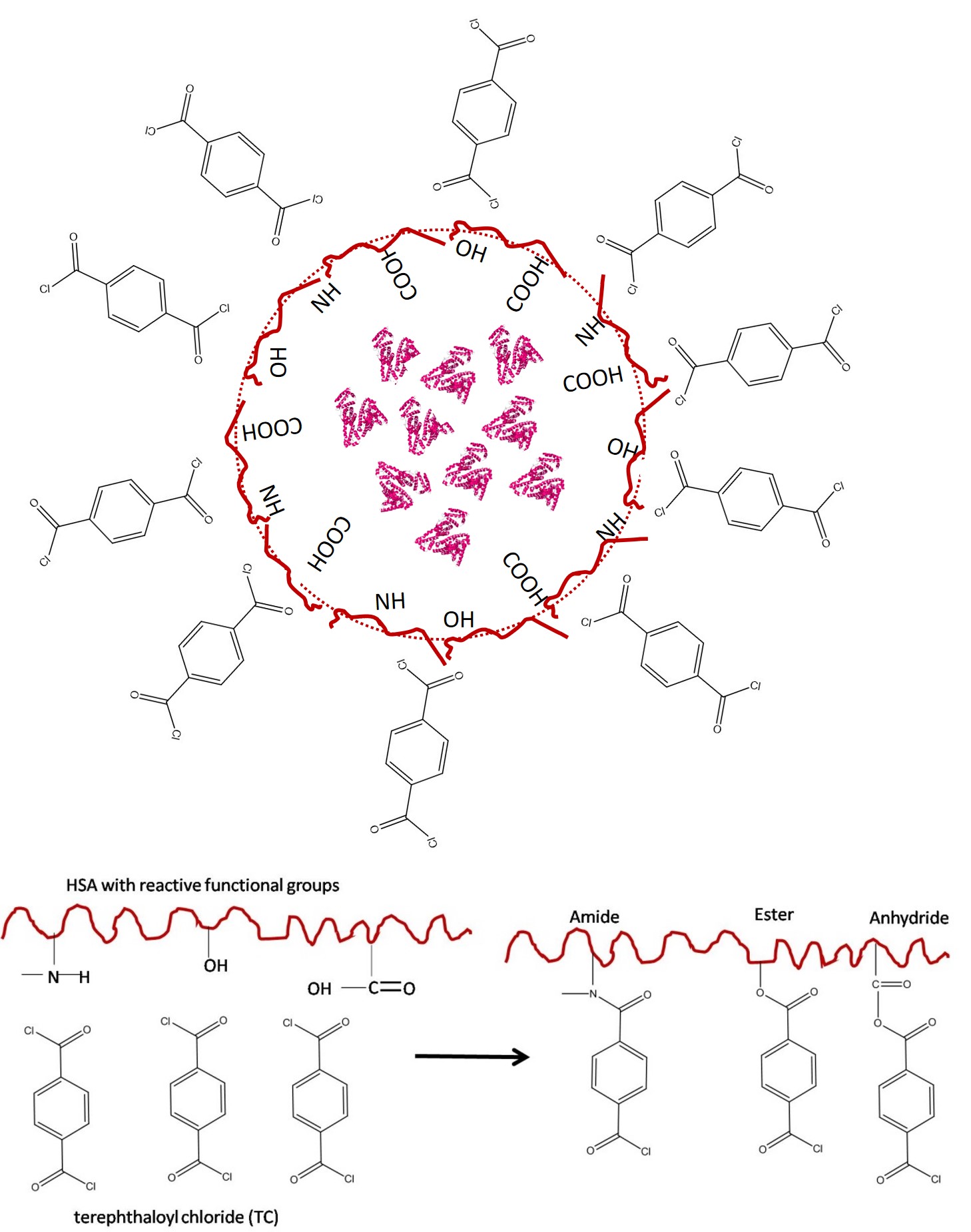}
 	\caption{}
 	\end{subfigure}
 	~
 	\begin{subfigure}{0.7\linewidth}
 	\includegraphics[width=1\linewidth]{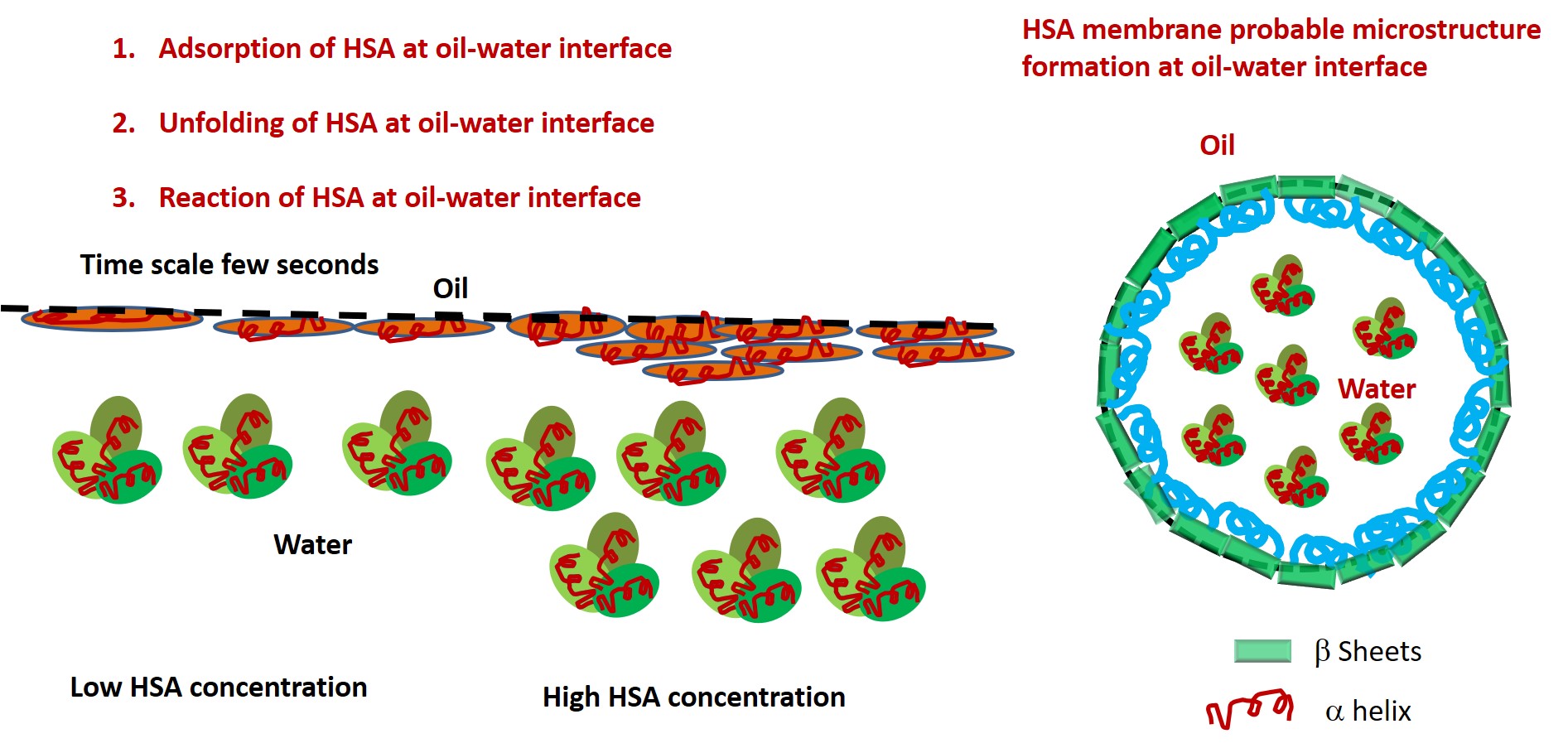}
 	\caption{}
 	\end{subfigure}
 	\caption{Schematic representation of (a) reaction mechanism (b) membrane formation mechanism for HSA microcapsule (not to the scale)}
 	\label{fig:reaction_mechanism_hsa}
 \end{figure}
 
 \begin{figure}
	\centering
	\includegraphics[width=0.7\linewidth]{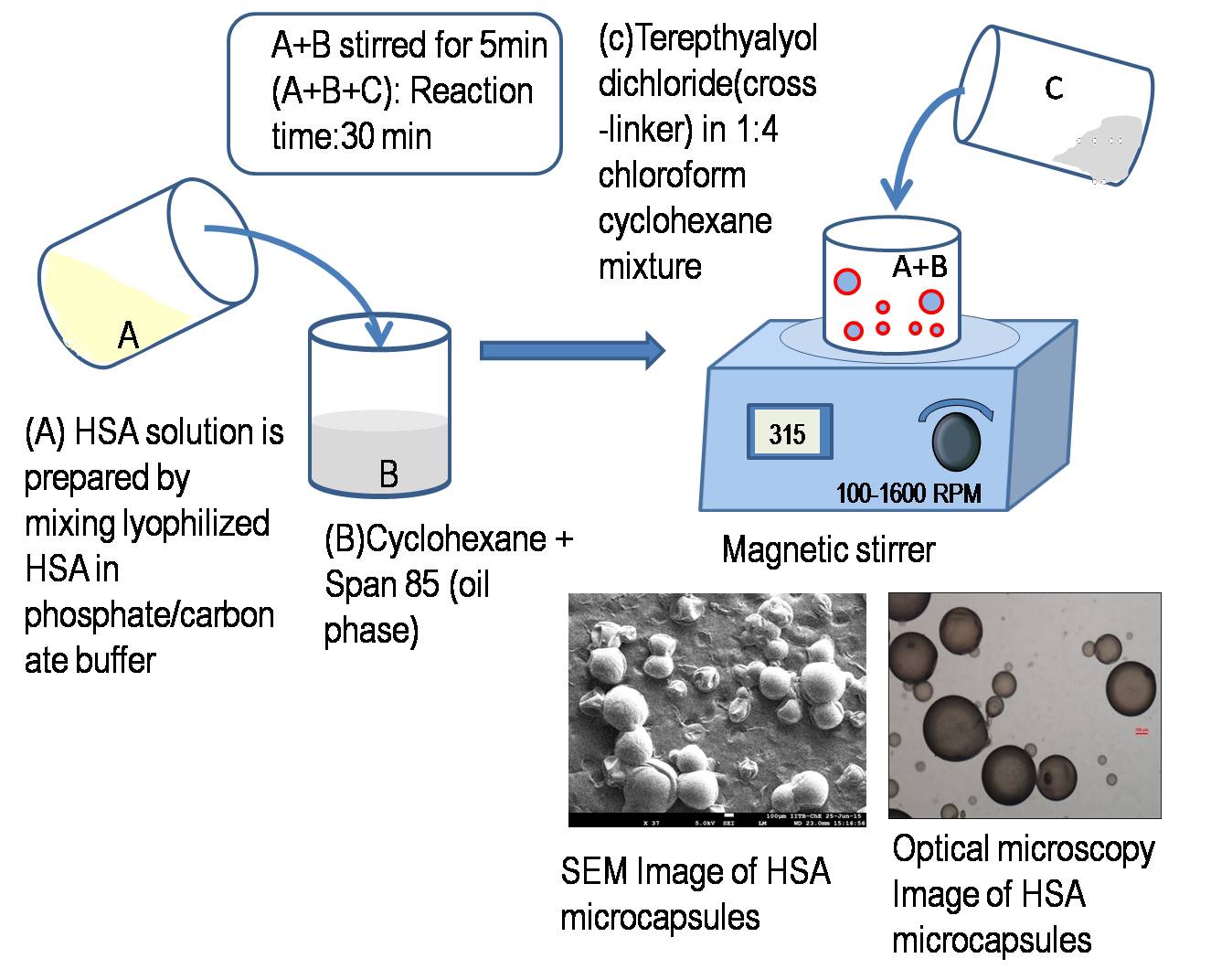}
	\caption{Methodology for the synthesis of human serum albumin microcapsules}
	\label{fig:process_of_making_microcapsules}
\end{figure}

\subsection{Methodology for synthesis of HSA capsules}

Figure \ref{fig:reaction_mechanism_hsa} shows the probable reaction mechanism for the formation of the membrane of HSA microcapsules.  HSA from the aqueous phase adsorbs at the oil-water interface and subsequently unfolding of proteins can occur, making the buried functional groups, amine, ester, and hydroxylic of proteins present in the aqueous phase available for reaction with the chloride group of the crosslinker terephthaloyldichloride present in the oil phase, to form amide, ester, and anhydride bonds. Adsorption and unfolding phenomenon of a protein depend on the reaction parameters, such as protein concentration and pH. At a high concentration of HSA, there might be competition between adsorption and the unfolding process resulting in the multilayer film formation. The hydrophobic $\beta$-sheets would mostly be on the oil side and the $\alpha$-helix  would be on the aqueous side of the membrane.

The HSA microcapsules were synthesized by the conventional method of interfacial polymerization of HSA using the emulsification technique with terephtyalyol chloride as the crosslinking agent \citep{levy1991fourier} as shown in Figure \ref{fig:process_of_making_microcapsules}. The lyophilized HSA  was dissolved in the phosphate (pH 7.4) or carbonate (pH 9.2) buffer to make solutions of different [10, 20, and 30 percent (weight/volume)] concentrations of HSA. Note, \% w/v= 100 gms of solute/100 ml of solution. Thus, a 10\% HSA solution had 0.1 gms of HSA/100 ml of solution, or 100 mg/ml of HSA. Around $100 \mu l$ of the HSA solution was then added to cyclohexane containing 5 percent (volume percent) span 85 and stirred at a constant speed of about 600 rpm for 5 min, to produce an emulsion of aqueous droplets containing HSA dispersed in cyclohexane. The cross-linker terepthylyoldichloride (0.25 percent weight/volume), dissolved in (1:4) chloroform cyclohexane mixture, was then added to this emulsion, and the polymerization reaction was stopped after 15 min by dilution of the reaction media. The capsules were washed gently thrice with an organic phase to remove the excess crosslinker. 
 
In the present work, we use the notations $HSA_{10}$, $HSA_{20}$, and $HSA_{30}$ to represent capsules synthesized with PBS buffer at pH 7.4 with concentrations of  10, 20, and 30 percent HSA respectively. A HSA capsule synthesized with carbonate buffer at pH 9.2 for 20 percent HSA is represented as $HSA_{20}, pH 9.2$. Poly-dispersed microcapsules were formed by this method in the size range of 10 $\mu m $ to $400$ $\mu m$.  Any one of these capsules from the emulsion was then used for more detailed interfacial study and the size of such a capsule was estimated from images obtained using light microscopy. The size distribution of the suspension of capsules (See Figure \ref{fig:size_distribution} provided in Supplementary Information) was studied using instrument particle size analyzer (Horiba LA 960). It should be noted here that for size measurement using the particle size analyzer, the capsules were resuspended in water, by removing the organic solvent using excess water. Thus the size of the capsules obtained using the Horiba (LA 960) are only indicative of the actual water-in-oil, crosslinked capsules, synthesized using the water-in-oil emulsion (For details see Supplementary Information).

\subsection{Scanning electron microscopy (SEM) studies for HSA capsules}

 The morphology of microcapsules was studied using the scanning electron microscopy (SEM) technique. Freshly synthesized capsules were placed on an aluminum foil using the drop-casting method and the organic phase was evaporated at room temperature to get the dry sample. The samples were coated with platinum for 60 s. Imaging was carried out in the dry mode at 5 kV with various magnifications ranging from 45X to 50,000X. 
\subsection{FTIR studies on capsules}

 The seccondary structure of proteins has been widely studied using the FTIR spectroscopic technique, especially the Amide-I, Amide-II band \citep{singh1993fourier,usoltsev2020ftir,yang2015obtaining,usoltsev2020ftir}. The microcapsules synthesized using the method described, were centrifuged and washed thrice with the organic phase and thrice with deionized water, and finally suspended in deionized water. The lyophilized microcapsules were then studied using the standard procedure for a wavelength range of 6250 to 5882 nm using the FTIR (Bruker, Germany, 3000 Hyperion Microscope with Vertex 80 FTIR System) instrument. 

\subsection{Experimental set-up for electrodeformation study:}

 \begin{figure}
 	\centering
 	\includegraphics[width=0.9\linewidth]{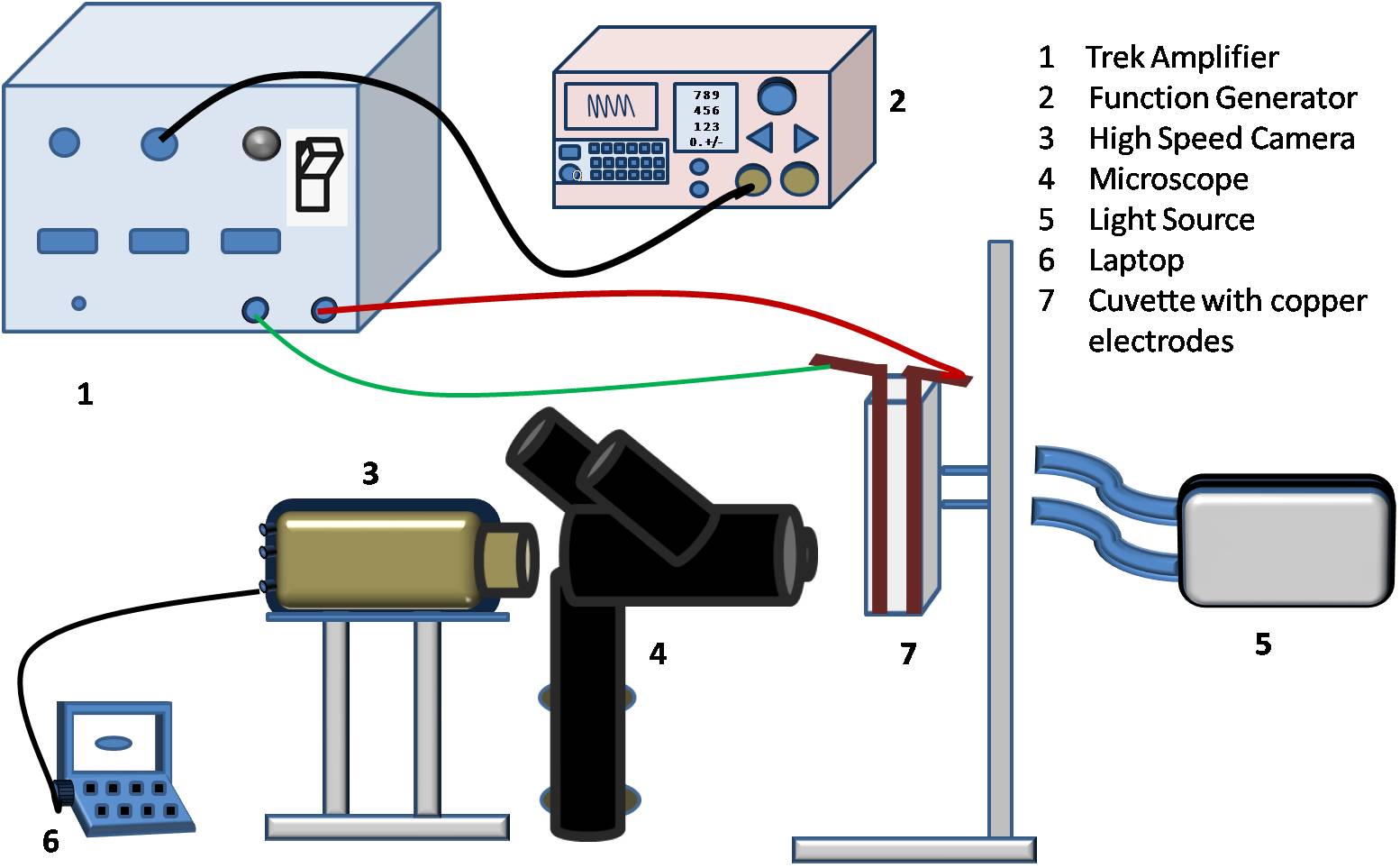}
 	\caption{Electrodeformation experimental set-up}
 	\label{fig:ehd-setup2}
 \end{figure}
 
 The experimental set-up (see Figure \ref{fig:ehd-setup2}) consisted of two copper electrodes of $45$ mm in height, 1 mm in thickness, and 1 cm wide, attached to the inner sidewalls of a plastic cuvette. The distance between the copper electrodes was kept as $4$ mm. A required voltage difference was applied across the copper electrodes with the help of a function generator (33220A Agilent Technologies Pvt. Ltd., USA) that had a voltage range of $ 0-20$ V. The frequency of the ac voltage used in this work, was in the range of (1Hz-1kHz). The function generator was connected to a high voltage amplifier TREK (model 20/20C-HS) of fixed gain $2000 $ $V/V $ to generate the desired amplified voltage. 
 
 For the mechanical characterization, a single microcapsule was manually placed in the cuvette containing 350 cSt silicone oil with the help of a micropipette. As the electric field was applied, the capsule deformed into an ellipsoidal shape in the direction of the field. The complete electrodeformation video (when the field is switched on and also after switching off the field) was recorded with the help of a high-speed videography (by Phantom V12 camera), attached to a  stereo-microscope (Leica z16AP0) at $200-70000$ frames per second. A light source  (Nikon) was used for illumination and image analysis was performed using the ImageJ software (https://imagej.nih.gov/ij/). 
 
 The deformation of the microcapsules was estimated by defining a modified Taylor parameter. 
 \begin{equation}
 	D =\dfrac{((l/L)-(b/B))}{((l/L)+(b/B))}
 	\label{deformaton  def eq_HSA}
 \end{equation} 
 Where L, B, l, and b  are the initial major and minor axes of the original undeformed and deformed capsule before and after the application of the electric field respectively. The surface Young's modulus is calculated using  Eq.\ref{Ca D0 theory eq_HSA} (See \citep{karyappa2014deformation} for details), which relates electric capillary number ($Ca_e$) to deformation (D), as expressed below.
 \begin{equation}
 	D =\dfrac{16}{45}Ca_e  
 	\label{Ca D0 theory eq_HSA} 
 \end{equation}
 \begin{equation}
 	Ca_e =\dfrac{a E_o^{2}\epsilon\epsilon_{o}}{E_{s}}
 	\label{Eq.:Ca_HSA }
 \end{equation}
 Here, $E_o$ is the RMS value of the applied electric field, a is the radius of a capsule, $\epsilon_o$ is permittivity of free space, $\epsilon$ is the dielectric constant of the medium and $E_s$ is the surface Young's modulus of a membrane (N/m). 
 
 \subsection{Viscoelastic model for the study of rheological properties of microcapsules} 
 
 \begin{figure}
 	\centering
 	\begin{subfigure}{0.4\linewidth}
 		\includegraphics[width=1\linewidth]{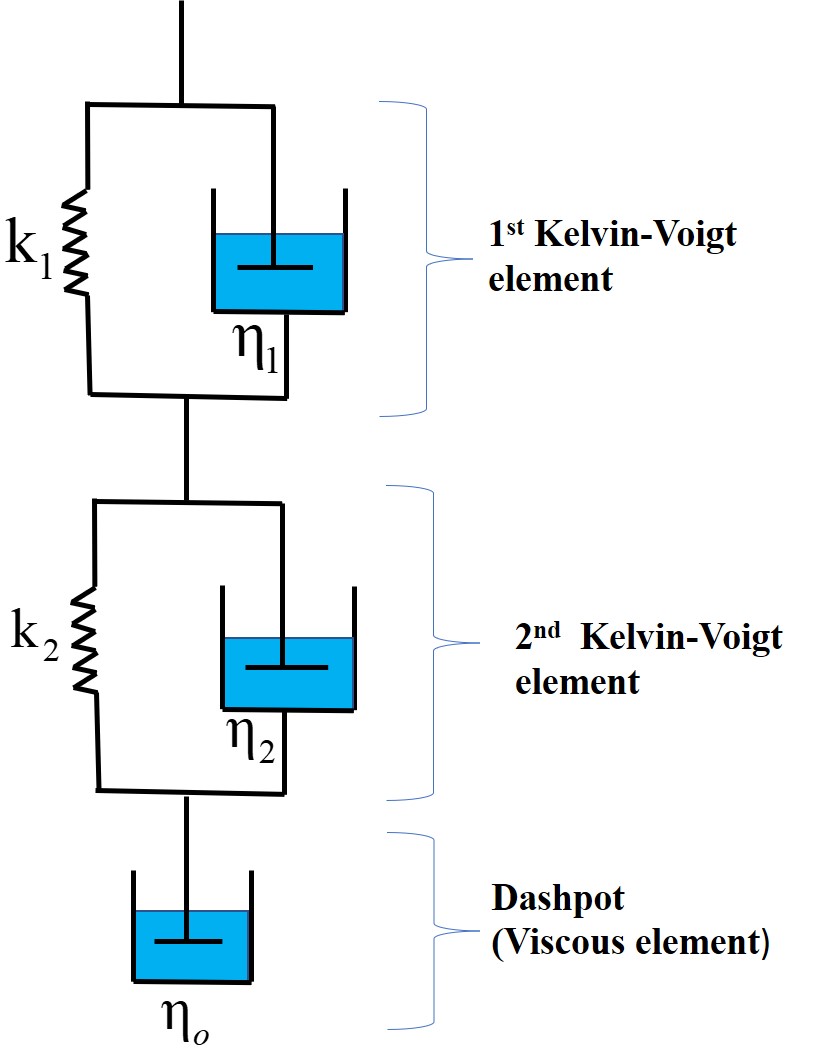}
 		\caption{}
 	\end{subfigure}
 	~
 	\begin{subfigure}{0.3\linewidth}
 		\includegraphics[width=1\linewidth]{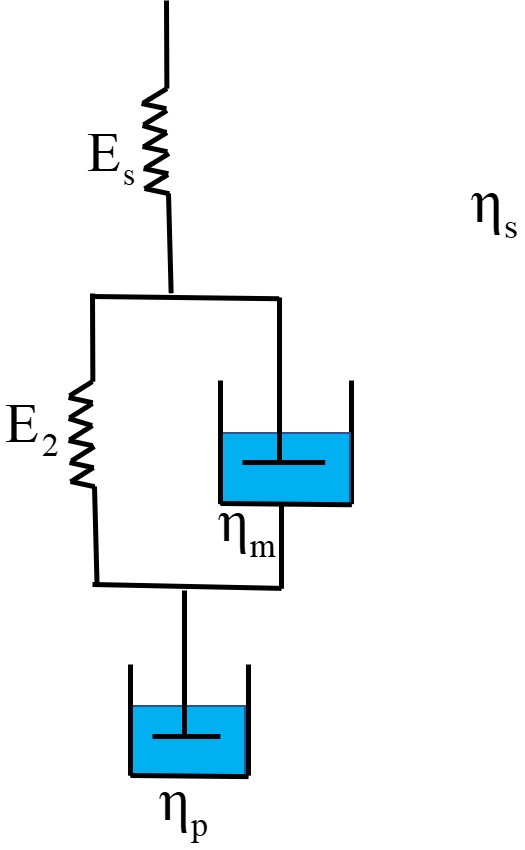}
 		\caption{}
 	\end{subfigure}
 	\caption{(a) Schematic of spring-dashpot visco-elastic model (b) Corresponding schematic of stress-strain constitutive equation for elasto-electro-hydrodynami model}
 	\label{fig:Schematic_viscoelastic_model}
\end{figure}
 
In the present work, we describe the rheological characteristics of the human serum albumin capsules using the five-element Burger model. The five-element viscoelastic model (Figure \ref{fig:Schematic_viscoelastic_model}) consists of two Kelvin-Voigt elements and one dash-pot arranged in series used for fitting the deformation versus time data for creep experiments. The complete response, that is deformation with respect to time, for a step-stress is proportional to $a \epsilon\epsilon_e E_o^2$  for $t<t_o$ is given by,
 \begin{equation}
 	D(t)=\left(\frac{1}{k_1} \left(1-e^{-t/\tau_1}\right)+\frac{1}{k_2}\left(1-e^{-t/\tau_2}\right)+\frac{t}{\eta_o} \right) a \epsilon \epsilon_o E_o^2
 	\label{Eq:model_eq_1}
 \end{equation}
 and for $t>t_o$
 
 \begin{equation}
 	D(t)=\left(\frac{1}{k_1} \left(e^{-(t-t_o)/\tau_1}-e^{-t/\tau_1}\right)+\frac{1}{k_2} \left(e^{-(t-t_o)/\tau_2}-e^{-t/\tau_2}\right)+\frac{t_o}{\eta_o} \right) a \epsilon\epsilon_o E_o^2
 	\label{Eq:model_eq_2}
 \end{equation}
 where $\tau_1=\eta_1/k_1$ and $\tau_2=k_2/\eta_2$.\\

Note that the elastic and viscous constants $k_1$ , $k_2$, $\eta_1$, $\eta_2$, in the spring-dashpot model, do not exactly represent membrane elasticity and viscosity. It is expected that $\eta_1$, $\eta_2$ should represent the solution viscosity $ a \mu_e$ shown in Figure    \ref{fig:Schematic_viscoelastic_model} (b)) and membrane viscosity $\eta_m$, respectively, with $ k_1$, $ k_2$, representing the membrane surface elasticity. A more accurate way is to have these spring-dashpot constants expressed as equivalent frame invariant stress-strain measures, which when solved in the spherical coordinate system, yield non-trivial geometry and property dependent parameters (such as viscosity ratio etc).  
 
 Therefore an viscoelasto-electrohydrodynamic model  developed in our earlier work \citep{puri2022study} which relates the parameters of the spring dash-pot model with the elasticity, membrane viscosity and solution viscosity as mentioned earlier, that is $\eta_1 \sim \mu_e a$ and $\eta_2 \sim \eta_m$, while $ k_1 \sim E_s$ and $k_2 \sim E_y$  is used. The interfacial stress-strain relationship for this 4-element linearised elastic model is given by, 
 \begin{gather}
 	\tau_{\theta \theta}^{el}+2 \frac{E_s \eta_p+ E_s \eta_m+E_2 \eta_p}{E_s E_2} \frac{\tau_{\theta\theta}^{el}}{d t}+\frac{\eta_m \eta_p}{E_s E_2} \frac{d^2 \tau_{\theta\theta}^{el}}{d t^2}= \eta_p \frac{d \epsilon_{\theta\theta}}{d t}+\frac{\eta_m \eta_p}{E_2} \frac{d^2 \epsilon_{\theta\theta}}{d t^2} \\
 	\tau_{\phi \phi}^{el}+2 \frac{E_s \eta_p+E_s \eta_m+E_2 \eta_p}{E_s E_2} \frac{\tau_{\phi\phi}^{el}}{d t}+\frac{\eta_m \eta_p}{E_s E_2} \frac{d^2 \tau_{\phi\phi}^{el}}{d t^2}= \eta_p \frac{d \epsilon_{\phi\phi}}{d t}+\frac{\eta_m \eta_p}{E_2} \frac{d^2 \epsilon_{\phi\phi}}{d t^2} 
 \end{gather}
 where subscripts "$\phi\phi$" and "$\theta \theta$" represent the azimuthal and meridional quantities, and $\epsilon_{\theta,\theta},\epsilon_{\phi,\phi}$ represent the linearised strain. This coupled with the electric and hydrodynamic stresses yields a relationship between the capsule deformation and the elastic, electric, and hydrodynamic parameters. The detailed model derivation is described in the Supplementary Information of our previous work \cite{puri2022study}.
 
 The equations (\ref{Eq:model_eq_1}, \ref{Eq:model_eq_2}) were fitted (Figure \ref{fig:creep_fit}) using the MATLB2019a and the parameters $ k_1, k_2,\eta_1,\eta_2$, and $\eta_o$ were estimated. The estimated fitting parameters are related to the elasticity, membrane viscosity, solution viscosity by the detailed viscoelasto-electrohydrodynamic model as mentioned in the Eq. \ref{eq:parameter_relation}.  
 \begin{gather}\label{eq:parameter_relation}
 	{k_1}  =\frac{16 {E_{s}}}{45}\text{;} \,\,\,\,\, 
 	{k_2} =\frac{16 {E_{y}}}{45}\text{;} \,\,\,\,\,
 	{\eta_{1}}=\frac{64 a{\mu_{e}}}{105}\,\,\,\,\,
 	{\eta_{2}}=\frac{32{\eta_{m}}}{45}\text{;}\,\,\,\,\, 
 	{\eta_{o} }=\frac{32{\eta_{p}}}{45} \text{;} 
 \end{gather}

 \subsection{Viscoelasto-electrohydrodynamic model for oscillatory test}
 
 The effect of frequency on the viscoelastic properties such as elasticity and membrane viscosity was studied by applying an oscillatory (ac) electric field, of varying frequencies, to the HSA microcapsules. The equations can be solved using the Laplace Transform technique when a sinusoidal electric field is applied. The equivalence between the complete hydrodynamic model and the spring-dashpot Burgers model is again found to exist similar to that of step stress case and the response is given by  
 \begin{align}
 (	D/a \epsilon\epsilon_o E_o^2 )= \frac{1}{2{k_1}}+\frac{1}{2 {k_{2}}}+\frac{t}{2 {\eta_3}} -\frac{2 {\tau_{1}}^2 \omega ^2 e^{-\frac{t}{{\tau_1}}}}{4 {k_1} {\tau_1}^2 \omega ^2+{k_1}}-\frac{2 {\tau_2}^2 \omega ^2 e^{-\frac{t}{{\tau_2}}}}{4{k_2} {\tau_2}^2 \omega ^2+{k_2}}- \nonumber  \\
 	\left (\frac{1}{2 \left(4 {k_1} {\tau_1}^2 \omega ^2+{k_1}\right)}+\frac{1}{2 \left(4 {k_2} {\tau_2}^2 \omega ^2+{k_2}\right)}\right)\cos (2  \omega t ) + \nonumber \\
 	\left(-\frac{1}{4 {\eta_3} \omega }-\frac{{\tau_1} \omega }{4 {k_1} {\tau_1}^2 \omega ^2+{k_1}}-\frac{{\tau_2} \omega }{4 {k_2} {\tau_2}^2 \omega ^2+{k_2}}\right) \sin (2  \omega t)
 \end{align}
 where the relationship between the parameters of the spring dashpot Burgers model to that of the full viscoelasto-electrohydrodynamic model remains the same. 
 
 The above expression can be used to fit and estimate the $E_s,E_y,\eta_m$ and $\eta_p$ values from the deformation time series for each frequency.

 \subsection{Methodology for estimation of model parameters and confidence bounds}
  
  The variation of J(t) (ratio of deformation to electric stress) with time was fitted using the equations \ref{Eq:model_eq_1}, \ref{Eq:model_eq_2} and \ref{eq:parameter_relation} to estimate the model parameters using MATLAB R2019a and Fminsearchbnd \citep{d2012fminsearchbnd} method for each capsule.
  The rheological properties of HSA capsules was estimated by fitting the mean deformation of all capsules for estimating the rheological properties from the fitting of plotted mean deformation with respect to stress. The confidence bounds were estimated for the parameters estimated from mean deformation of all capsules.
We computed the confidence bounds for the estimated parameters using the F test \citep{namdeo2022palladium}.

 \section{Results and discussion:}

\subsection{Different time scales in the system} 

The formation of a cross-linked protein film at an oil-water interface proceeds through several steps at different rates. When proteins are dispersed in an aqueous solution, they can self-aggregate. The time scale of self-aggregation of HSA molecules to form oligomers seems to be of the order of few microseconds, indicating that the formation is instantaneous, and aggregate sizes of around a few hundred nanometers have been reported\cite{chubarov2020reversible}. The aggregated and non-aggregated protein molecules then migrate to the oil-water interface over time scales of $a^2/D \sim (300 \mu m)^2/(6 \times 10^{-11}) \sim 1500 s$. The time scale though could be shorter because of convection in the system on account of stirring as well as very high concentrations of HSA used in this work, which can effectively decrease intermolecular and interaggregate distance as well as their distance from the interface.

The second relevant time scale is that of adsorption of proteins on the surface. This is reported to be of the order of $1$ s \citep{yano2012kinetics}. This is followed by protein unfolding, at the oil-water interface, which is known to occur over time scales of few seconds to minutes. Finally the crosslinking is expected to be instantaneous due to very fast reaction. Further, growth of the interfacial film can occur due to diffusion-controlled reaction since the HSA present inside the drop phase needs to overcome the barrier of the polymeric membrane formed to react further with the cross-linker present on the oil side. The other time scale of relevance, which is the diffusion time of TC to arrive at the interface is expected to be very short.

 \subsection{Characterization of capsules using Fourier transform infrared spectroscopy (FTIR) technique:}

 \begin{figure}[!ht]
    \centering
	\includegraphics[width=1\linewidth]{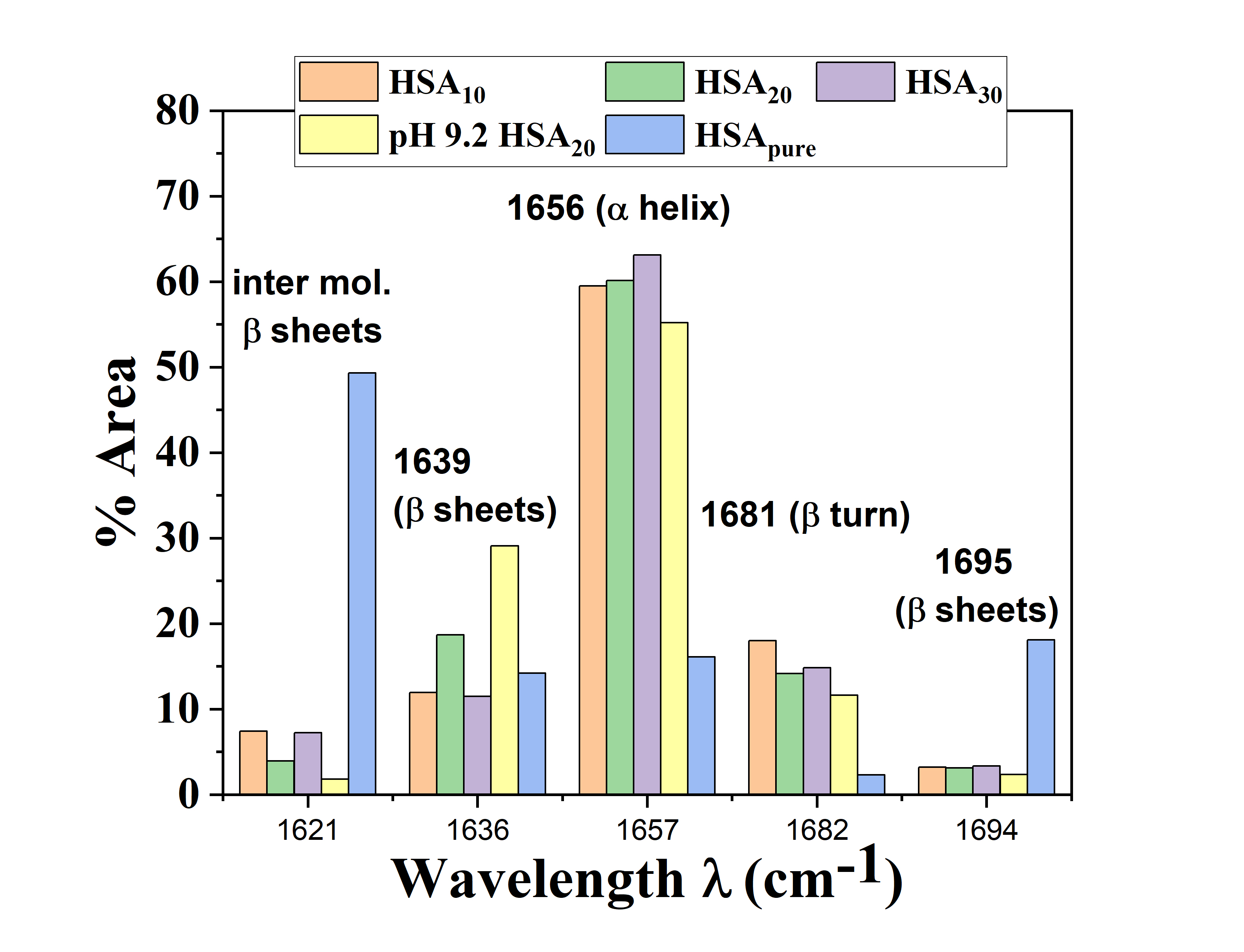}
    \caption{Comparison of secondary structure of original HSA and HSA capsules synthesized at different conditions}
    \label{fig:FTIR_area}
\end{figure}
 
 The amide -I band (1600–1700 cm–1), which has contributions from the $C=O$ stretching due to vibration of the amide group, the in-phase bending of the N-H bond and the stretching of the C-N bond, is the most sensitive spectral region representing changes in secondary structure of proteins \citep{yang2015obtaining}. Amide-II is derived mainly from in-plane N-H bending and the C-N stretching vibrations and is more complex than Amide-I. Although this band is conformationally sensitive, it has not been commonly used for describing changes in protein structures. The other amide bands are rarely used to study protein structure due to their complexity and dependence on the details of the force field, the nature of side chains, and hydrogen bonding\citep{yang2015obtaining}.

Figure \ref{fig:ftir} provided in Supplementary Information shows FTIR spectra in full-range for pure HSA and cross-linked HSA microcapsules studied in the present work. We analyzed the Amide-I band in detail to understand the microstructure of the membrane of the microcapsules using quantitative analysis of the Amide-I band contour, employing curve fitting, second derivative, and Fourier self-deconvolution methods. A Gaussian curve-fitting was conducted on deconvoluted spectra using OriginPro 2019b software. The peaks obtained by deconvolution of the amide-I band are assigned to specific types of secondary structure based on the known correlations between different secondary structures ($\alpha$ helix, $\beta$ sheets, random coil) of proteins and their FTIR spectra \citep{levy1991fourier,litvinov2012alpha,singh1993fourier}. 

From Figure \ref{fig:ftir} (See Supplementary Information), it is clear that there is a significant difference in the spectral peaks for uncrosslinked HSA in aqueous solution and cross-linked capsules. The deconvoluted Amide-I bandwidth spectra fitted for different cases of capsules as well as that of lyophoilized uncrosslinked HSA is shown in Figure \ref{fig:ftir_o} and \ref{fig:FTIR_fit} provided in Supplementary Information. Figure \ref{fig:FTIR_area}  shows percentages of different elemental structures such as $\alpha$-helix, $\beta$-sheets and random coils present in the capsules for different cases considered in the present study. It is important to note here that although the aqueous HSA molecules contain no $\beta$-sheets and have a completely helical structure, the lyophilized HSA used in our FTIR study showed a significant percentage of $\beta$-sheets. Infact, Figure \ref{fig:FTIR_area} indicates that the pure lyophilized HSA contains more $\beta$-sheets than the crosslinked HSA capsules. It is known that lyophilization of protein leads to unfolding and changes in their secondary structure, which upon rehydration in an aqueous buffer, might be again refolded entirely or partially \citep{lin2000stability}. 

 The capsules synthesized with pH 9.4 buffer showed more $\beta$-sheet formation than the capsules synthesized with PBS 7.4 buffer. This could be attributed to greater disruption of the native structure at higher pH due to charge repulsion. It is should be noted that the FTIR study was done on lyophilized capsules with polydispersed capsules synthesized by the emulsion technique. Therefore the variation of $\beta $ sheets with concentration at pH 7.4 can not be accounted due to polydispersity in the size of capsules. The results of FTIR studies for different concentrations of HSA should be analyzed with caution. The FTIR studies were conducted on a collection of capsules of varying sizes, as obtained in the emulsion technique. As shown later in the studies, the radius of the capsule can affect the crosslinking significantly. The different percentages of $\alpha$-helix, $\beta$-sheets, and random coil in capsules  can result in different membrane surface morphologies, which ultimately affect their mechanical properties.
At low pH,  free amines participate in the reaction while at higher pH, the carboxylic groups can ionise and participate in anhdyride bond formation along with ester bonds formed by acylation of hydroxyl groups \citep{levy1991fourier}. The involvement of different functional groups in the crosslinking could affect the crosslinking degree and formation of secondary protein structure.

\subsection{Scanning electron microscopic results for membrane surface morphology }

\begin{figure}
 		\centering
 	\begin{subfigure}{0.7\linewidth}
 		\centering
 		\includegraphics[width=1\linewidth]{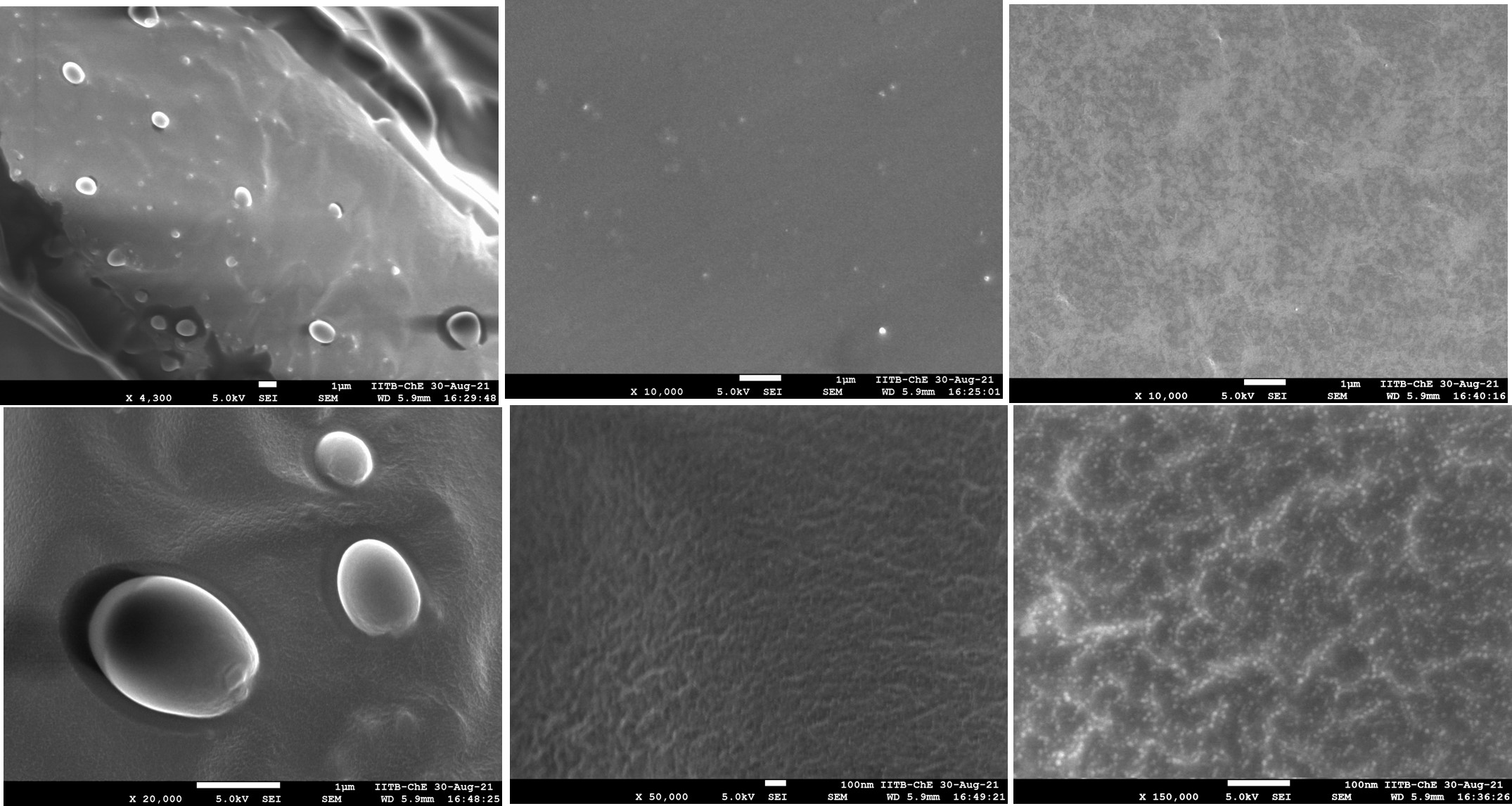}
 		\caption{}
 	\end{subfigure}
 	
 	\begin{subfigure}{0.7\linewidth}
 		\centering
 		\includegraphics[width=1\linewidth]{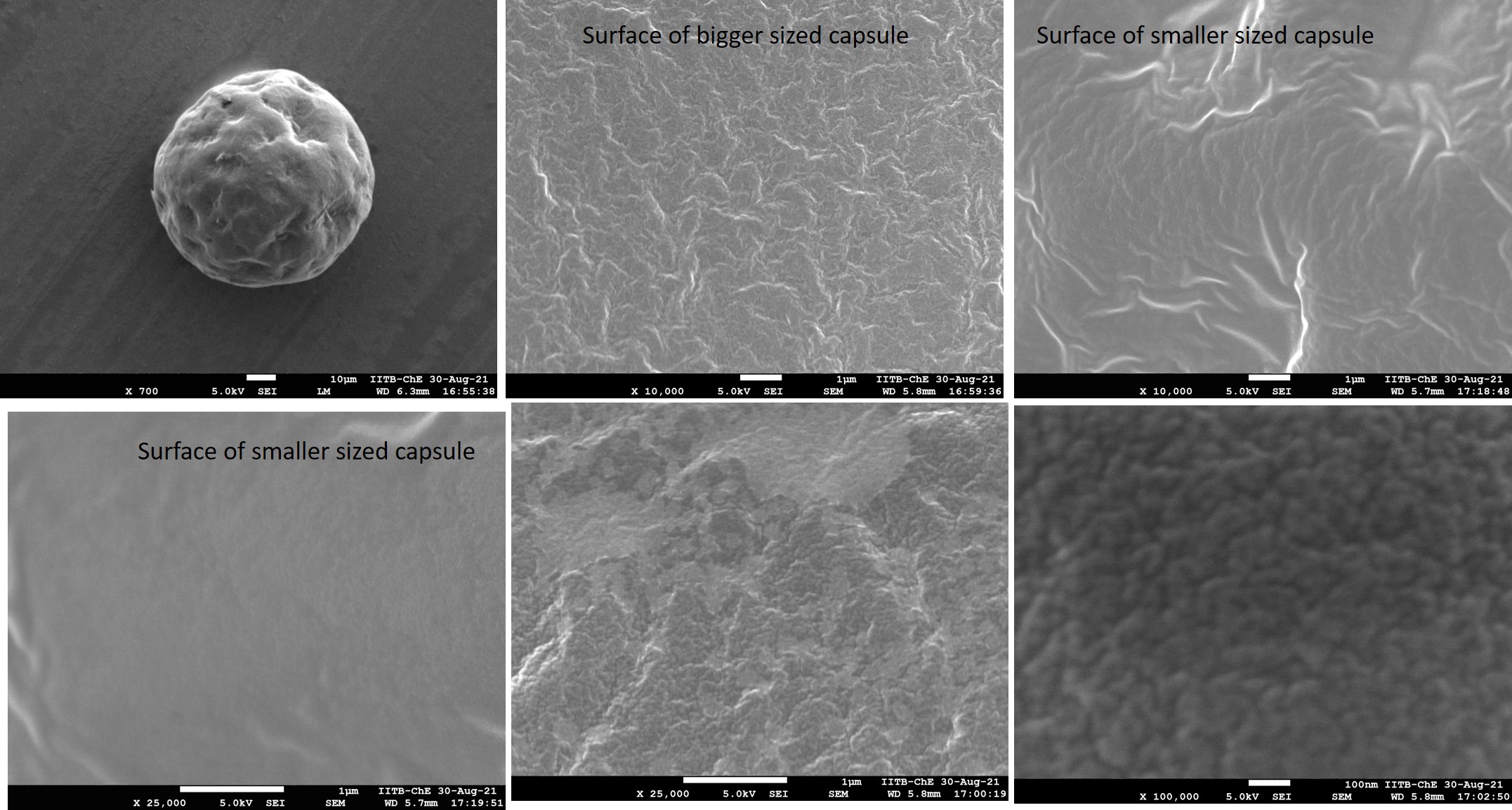}
 		\caption{}
 	\end{subfigure}
 	
 	\begin{subfigure}{0.7\linewidth}
 		\centering
 		\includegraphics[width=1\linewidth]{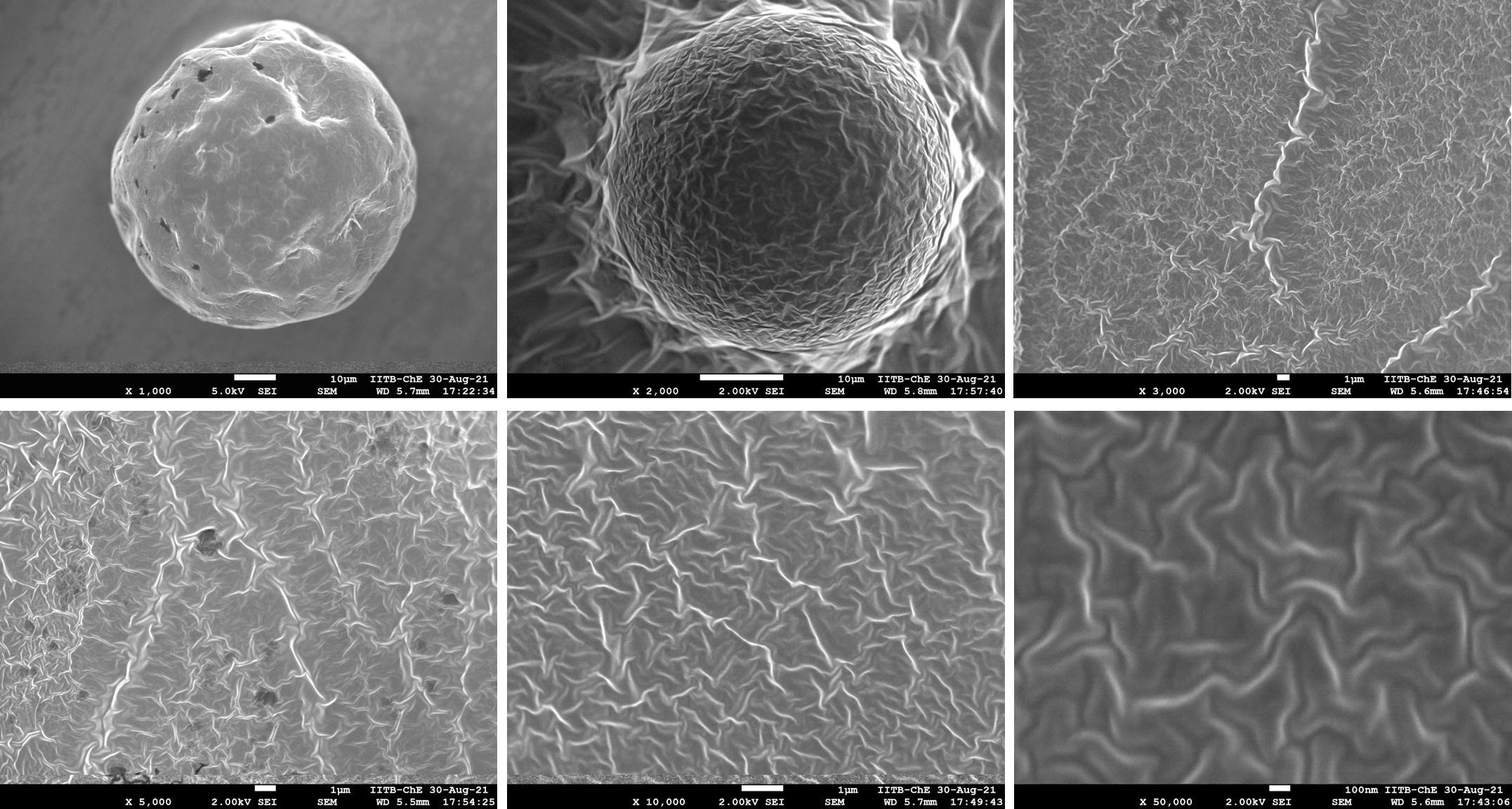}
 		\caption{}
 	\end{subfigure}
 	\caption{Scanning electron microscopy  images for capsules prepared with 15 min reaction time with phosphate pH 7.4 buffer for (a) $HSA_{10}$ (b) $HSA_{20}$ (c) $HSA_{30}$}
 	\label{fig:SEM_images_of_HSA_capsules}
   \end{figure}

 \begin{figure}
 	\centering
 		\begin{subfigure}{0.75\linewidth}
 	\includegraphics[width=1\linewidth]{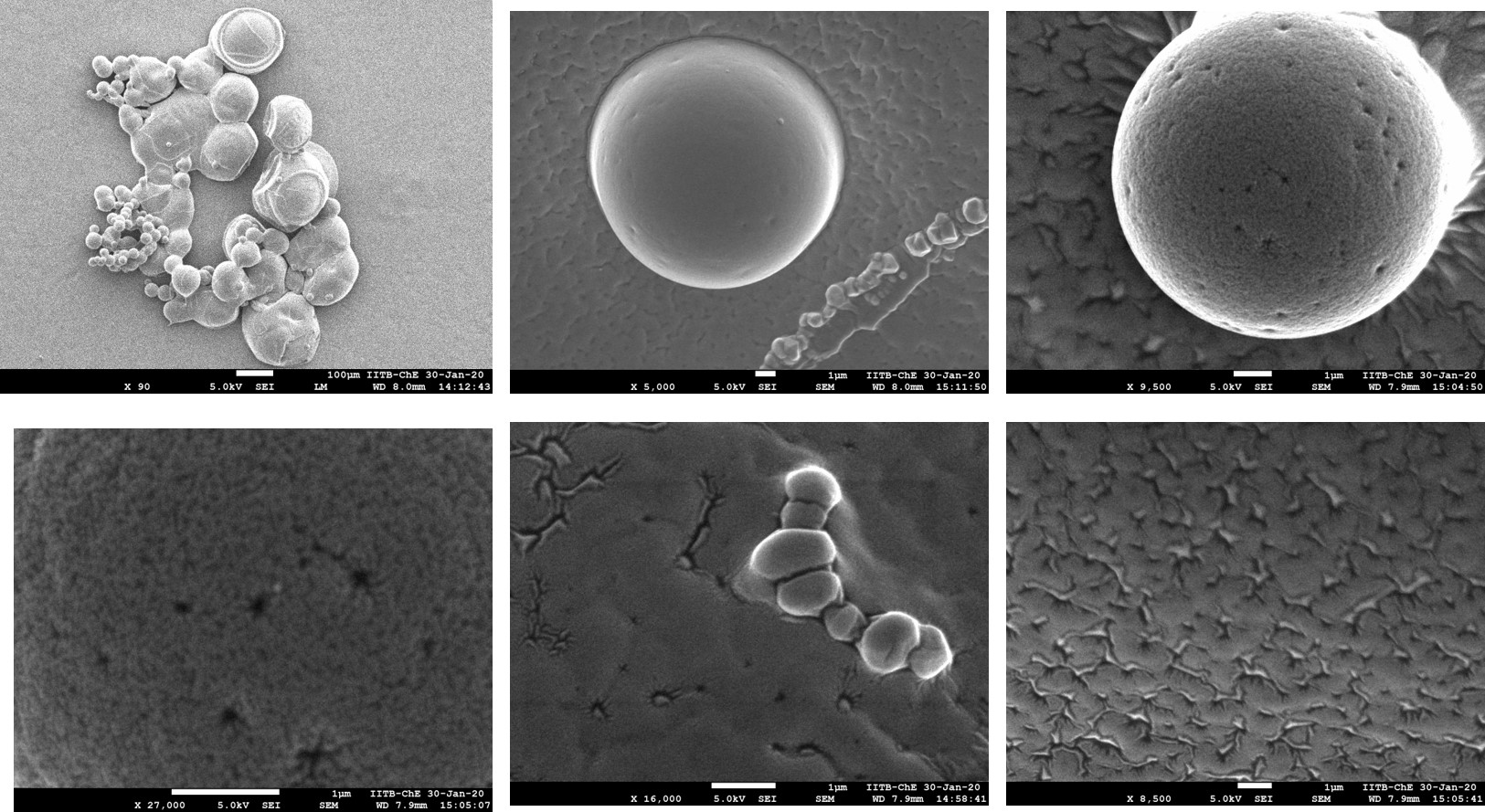}
 	\caption{}
 	\end{subfigure}
 	~
 	\begin{subfigure}{0.75\linewidth}
 	\includegraphics[width=1\linewidth]{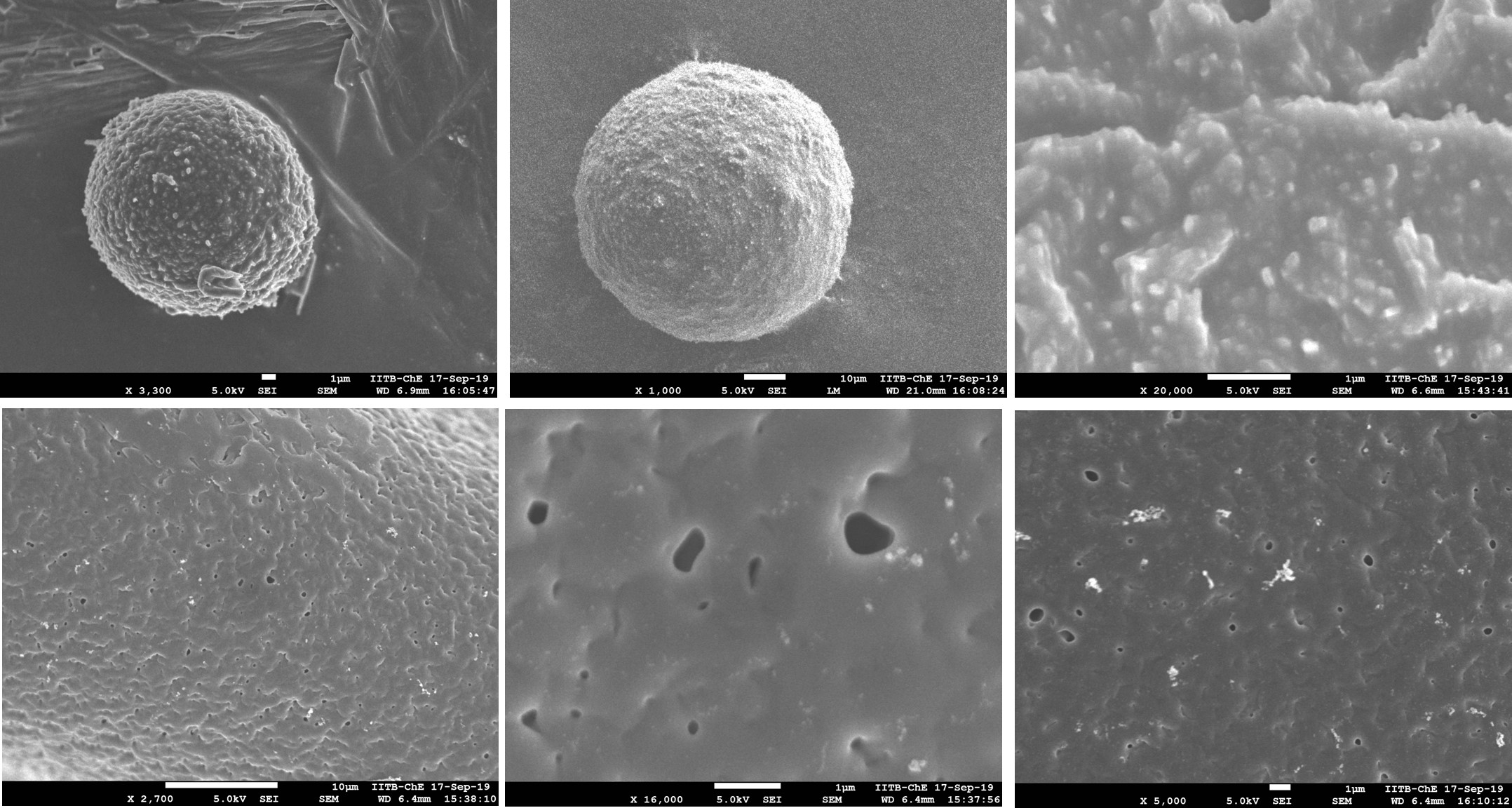}
 	\caption{}
 	\end{subfigure}
 	
 	\caption{Scanning electron microscopy images of HSA capsules synthesized with carbonate buffer at pH 9.2 (a) 10 percent HSA, 20 min reaction (b)15 percent HSA, 20 min reaction }
 		\label{fig:SEMcarbonate}
 \end{figure}

Figure \ref{fig:SEM_images_of_HSA_capsules} shows the surface morphology of the microcapsule membrane for different pH and concentrations of HSA studied in the present work using SEM images.
The different morphologies of capsules for different HSA concentrations could be due to the fact that the interfacial adsorption and unfolding phenomena depend upon the pH and concentration of HSA. These factors affect the cross-linking and thereby, the membrane morphology. The degree of cross-linking reaction and membrane microstructure depends upon the availability of buried functional groups of a protein at the interface. This in turn depends upon the protein conformational changes induced by a change in interface environments such as pH, HSA molecules per unit area, and the nature of the interface (aqueous, organic, air).

Figure \ref{fig:SEM_images_of_HSA_capsules} indicates that $HSA_{10}$ capsules show relatively smoother morphology than that for the other cases. SEM images showed that tiny microcapsules are adsorbed on the surface of bigger microcapsules in the case of $HSA_{10}$. The adsorbed smaller capsules show very smooth membrane surface morphology. This is possibly because the surface concentration of the protein $\Gamma = 4/3 \, a \, C $, where $C $ is the bulk HSA concentration for a capsule with size $a$. Thus reduction in size leads to lower surface concentration, lower reaction rates, and thereby lower crosslinking and a smoother surface, aided by lower surface concentration. 
 It should be mentioned here that the capsules synthesized with the emulsification method give a wide range of size distribution, and hence the smaller and larger microcapsules prepared with the same HSA concentration showed different morphology. Similar to $HSA_{10}$, smaller microcapsules with $HSA_{20}$ also showed relatively smoother membrane morphology than bigger microcapsules.

 In the case of $HSA_{20}$ (Figure \ref{fig:SEM_images_of_HSA_capsules} b), for the similar sized capsules, the capsules showed a rougher membrane surface than for the case of $HSA_{10}$, possibly due to higher concentration of HSA. The microcapsules synthesized with $30 \%$ HSA concentration showed a much different microstructure ( Figure \ref{fig:SEM_images_of_HSA_capsules} c) than for the other two cases ($HSA_{10}$ and $HSA_{20}$). For the case of $HSA_{30}$, the capsules showed a wavy folded structure. Few microcapsules of $HSA_{30}$ case showed pores on their membrane surfaces. The change in the morphology for $HSA_{30}$ could be attributed to the formation of larger oligomeric structures in the bulk at such high concentrations. These oligomeric structures can assemble at the interface and form porous, weakly crosslinked structures. 

Capsules synthesized with carbonate buffer at pH 9.2 showed a very rough, granular membrane surface with bigger pores ( Figure \ref{fig:SEMcarbonate}), identical to those seen for $HSA_{30}$. It is known that HSA changes its configuration from native form to a basic structure when pH changes from 7 to 9. This N-B transition; results in exposure of the
buried residues, leading to an increase in exchangeable
hydrogen and an increased bulk aggregation despite increased charge stability. This is supported by the greater conversion of $\alpha$-helix to $\beta$-sheets for pH 9.4 as seen in the FTIR studies. The assembly of these bulk oligomers at the interface can lead to a porous membrane morphology. Thus both high pH and high concentration can lead to increased hydrophobicity and formation of oligomers in the bulk aqueous solution. Further, we investigate the rheological properties of HSA capsules using the electrodeformation technique.

 \subsection{Justification for the five-element Burger model:} 
 
 The dynamic deformation of HSA capsules studied at different applied stresses shows an instantaneous deformation followed by slow time-dependent deformation. The instantaneous deformation can be related to the spring element $k_{1}$, associated with the solution viscosity giving a fast timescale. The $k_{2}$ and membrane viscosity ($\eta_2$), correspond to a slow creep response. The nonrecoverable creep deformation, is captured by the extra dashpot in series ($\eta_o$). From a microstructural viewpoint, the occurrence of the  elasticity can be attributed to the existence of intermolecular bondings. It has been suggested in the literature that the protein molecule exists on the surface in an unfolded structure, probably in the form of straightened $\beta$-configuration resulting in the instantaneous elastic response \citep{inokuchi1953rheology}. The extent of various secondary structural units such as $\alpha$-helix, $\beta$- sheets, and random coils and their conformational arrangements in cross-linked protein can result in the different rheological properties of the membrane. 

The literature does not present enough evidence on structure-rheology relationships for crosslinked proteins. For example, it is unclear from the literature if the proteins at the interface are folded or unfolded and the extent of conversion of $\alpha$ helices into $\beta$ sheets, as well as the percentage of intermolecular and intramolecular crosslinking. Based on the limited literature \citep{zhang2017self,brown1963physical,chan2020spider,debenedictis2019mechanical,choe2005elasticity,hsin2011molecular,buehler2008elasticity,ahmad2005guanidine} about the relation between different structural units of protein with the mechanical characteristics, for an interfacially crosslinked protein network, we propose a microstructure for the microcapsules as follows:
\begin{enumerate}
      \item The interface is laden with proteins adsorbed from the bulk, either as single molecules or as oligomers.
      \item The protein molecules then can unfold to varying degrees, typically, the domains I, II, and III could break the quaternary structures without significantly affecting their individual tertiary and secondary structures.
      \item Increased conversion of the $\alpha$-helix to the $\beta$-sheets indicates increased hydrophobicity. This could lead to a higher oligomerization in the bulk, or indicate a higher amount of unfolding at the interface, with hydrophobic parts residing in the oil phase.
      \item The unfolded structures could get crosslinked, most likely in the bulk, with the hydrophobic regions of the proteins anchored in the oil phase.
      
      \item The interfacial film is typically multi-layered.  The microstructure of the interfacial film and its crosslinking determines the rheological properties of the microcapsule. Multilayered and thick films are known to have  lower rheological properties \citep{puri2019study,jancar2008thickness}.
\end{enumerate}
One then expects two sources of elasticity. The anchored hydrophobic parts of the proteins in the oil phase yield an elastic response, which is resisted by the viscosity of the oil. This is expected to result in an instantaneous deformation, which could be attributed to the stretching of $\beta$-sheets. The subsequent creep could be because of the uncoiling of  $\alpha$-helices and random coil structure and the viscous resistance arising out of internal friction of the protein domains, giving the viscoelastic effect. The secondary structure can rearrange with the conversion of  $\alpha$-helix into $\beta$-sheets during the deformation, and hence the plastic deformation (unrecoverable creep) can be attributed to the irreversible changes in the secondary structure of a protein membrane.
 
 \subsection{Estimation of rheological characteristics of microcapsule membrane}
 
 \begin{figure}
      \centering
     \includegraphics[width=0.7\linewidth]{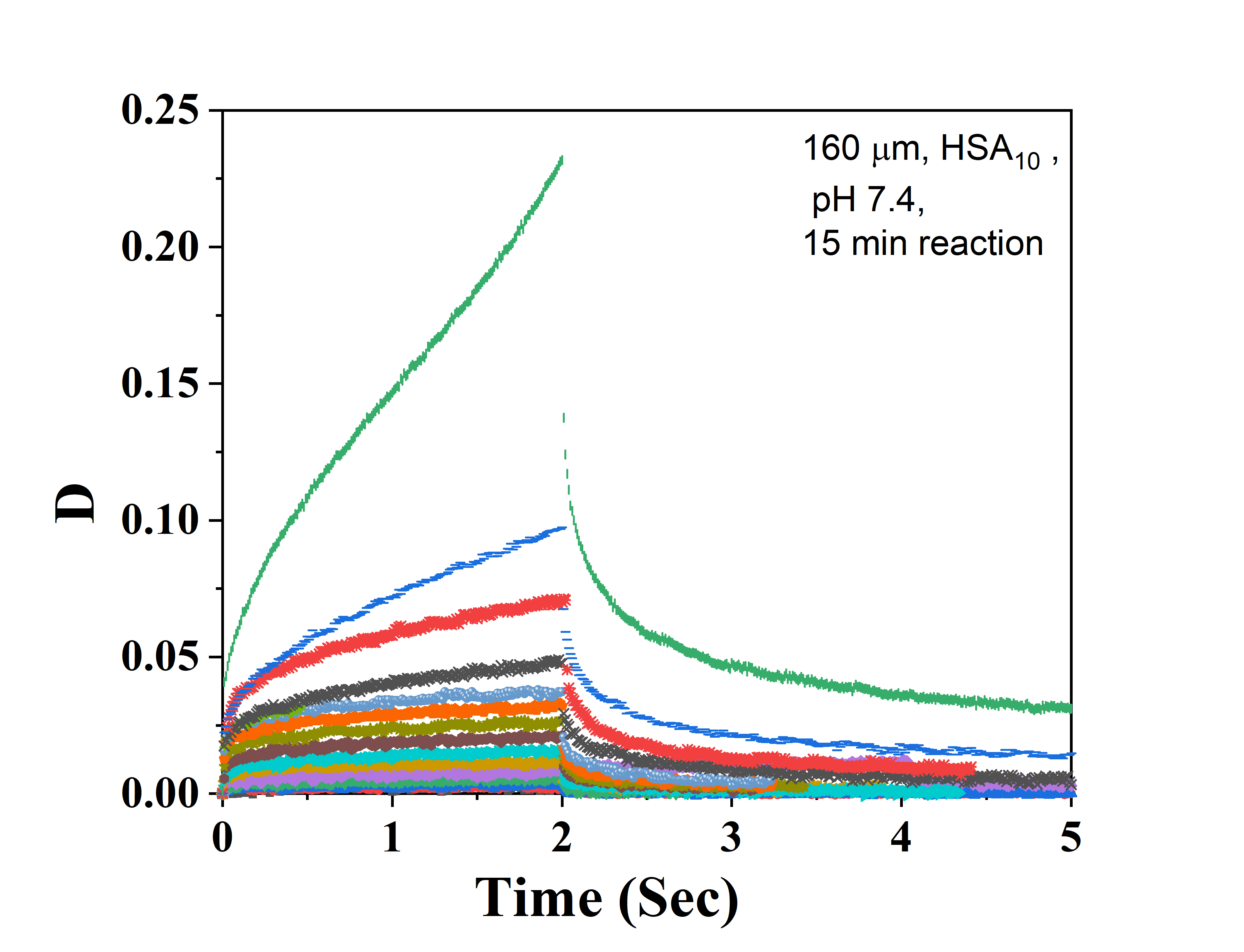}
      \caption{Dynamic deformation response HSA capsule for different electric stresses}
     \label{fig:D_vs_time_HSA}
  \end{figure}
  
  \begin{figure}
    
     \begin{subfigure}{0.45\linewidth}
          \includegraphics[width=1\linewidth]{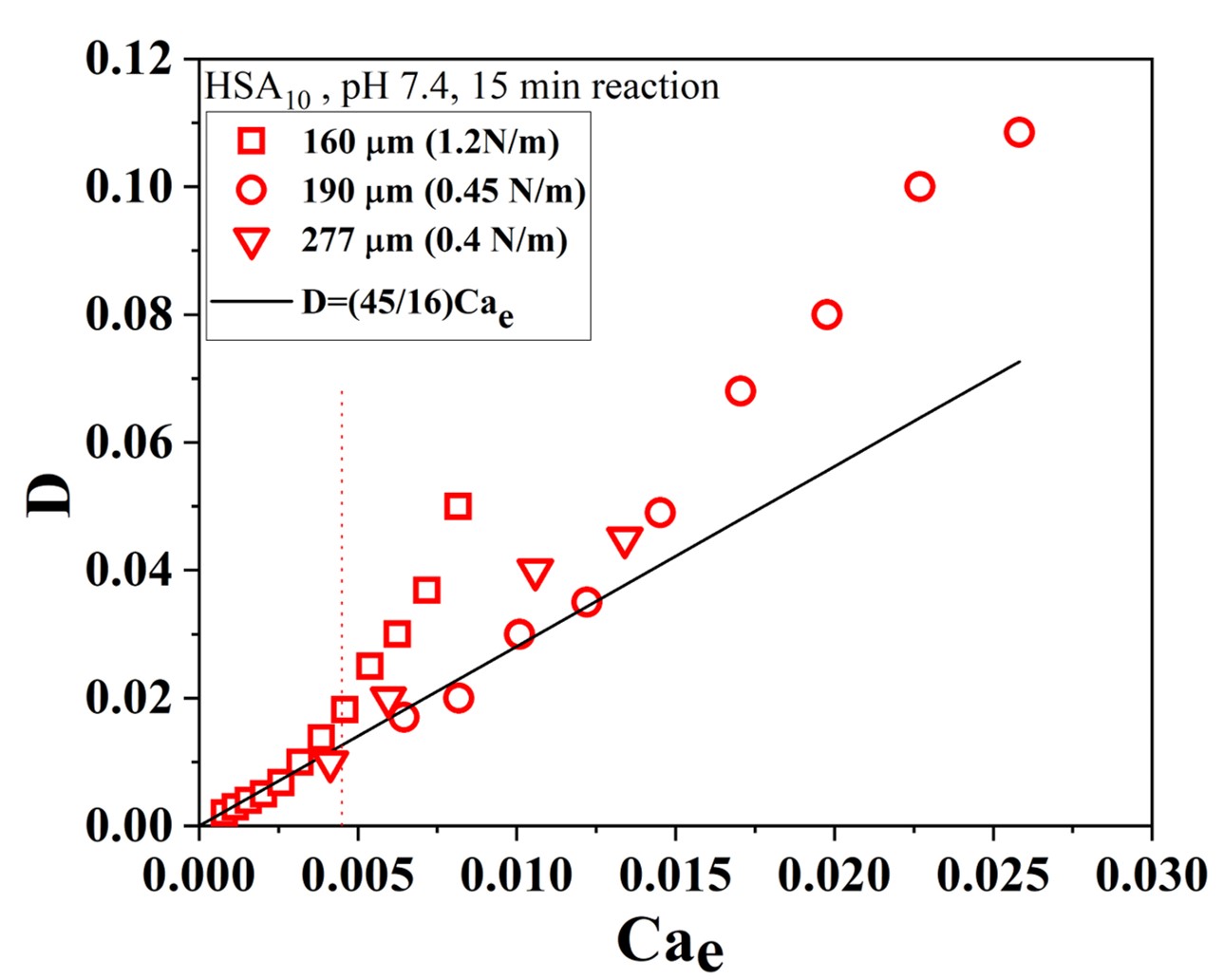}
     \caption{}
     \label{fig:D_vs_cae_set1}
     \end{subfigure}
     ~
     \begin{subfigure}{0.45\linewidth}
          \includegraphics[width=1\linewidth]{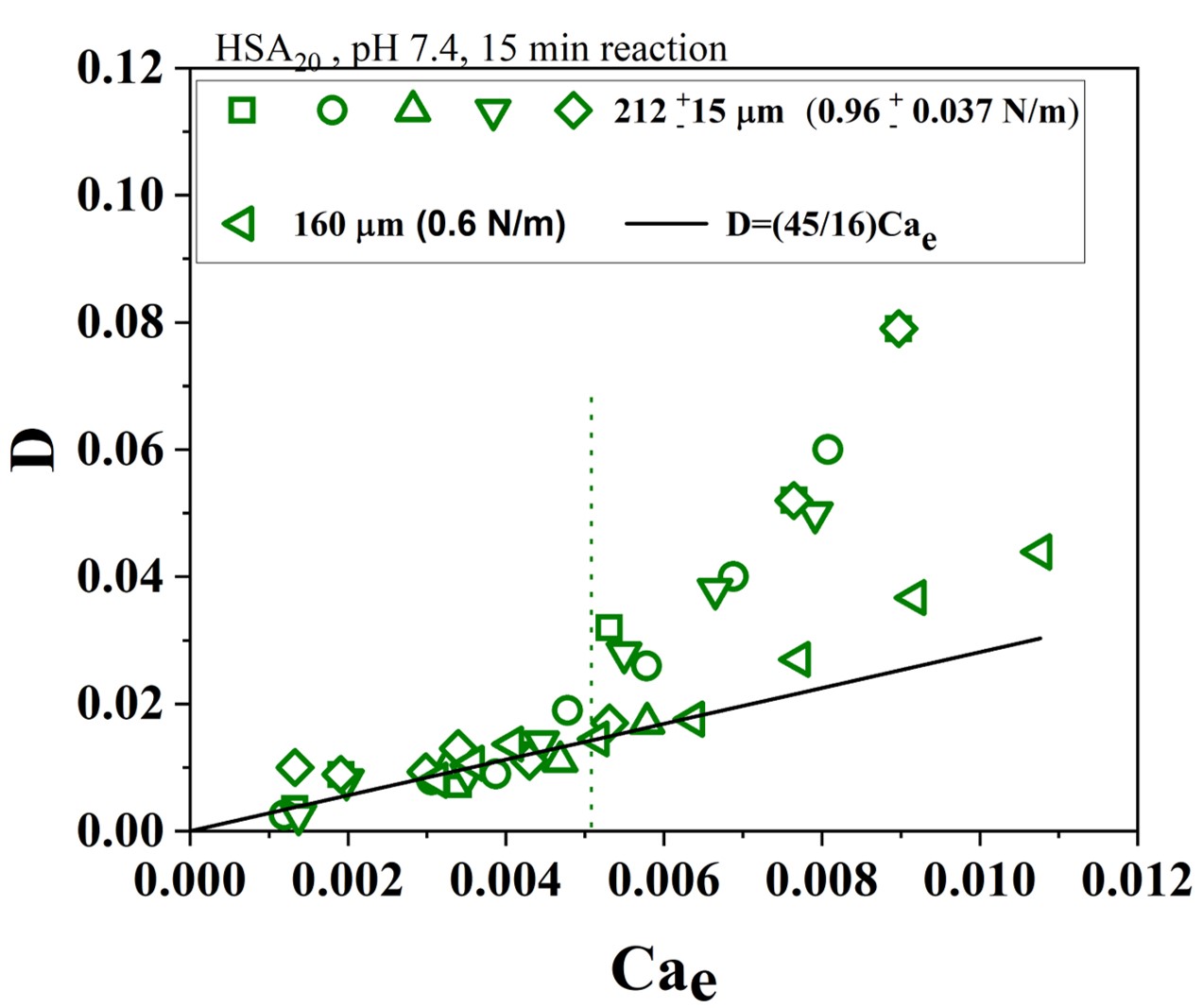}
     \caption{}
     \label{fig:D_vs_cae_set2}
     \end{subfigure}
     \caption{Deformation with respect to electric capillary number for HSA microcapsules synthesized with different HSA concentrations at pH 7.4 (a) $HSA_{10}$, (b)$HSA_{20}$}
     \label{fig:D_vs_Cae}
 \end{figure}
 
 \begin{figure}
 		\centering
 	\begin{subfigure}{0.45\linewidth}
 		\centering
 		\includegraphics[width=1\linewidth]{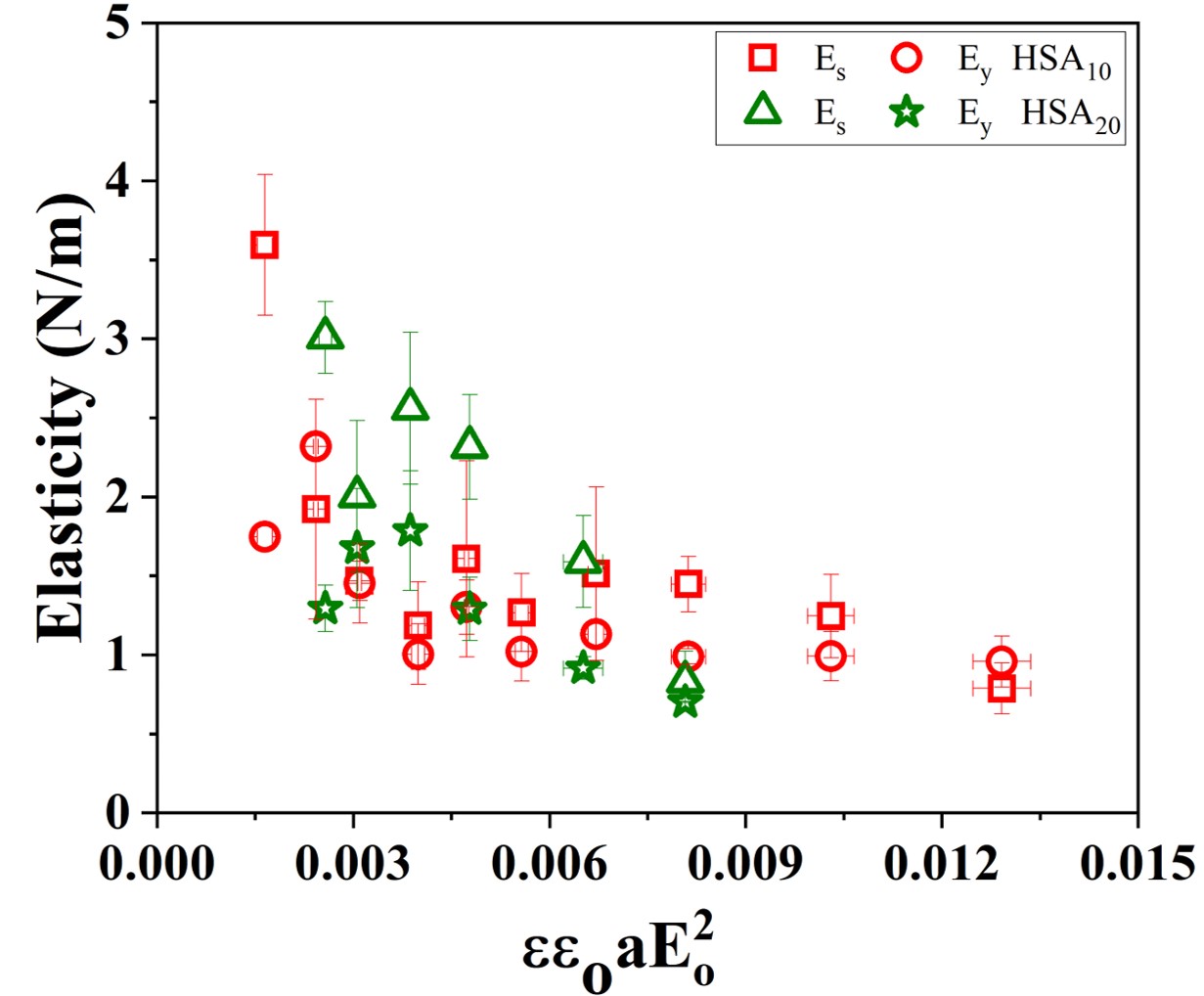}
 		\caption{}
 	\end{subfigure}
 	\begin{subfigure}{0.45\linewidth}
 		\centering
 		\includegraphics[width=1\linewidth]{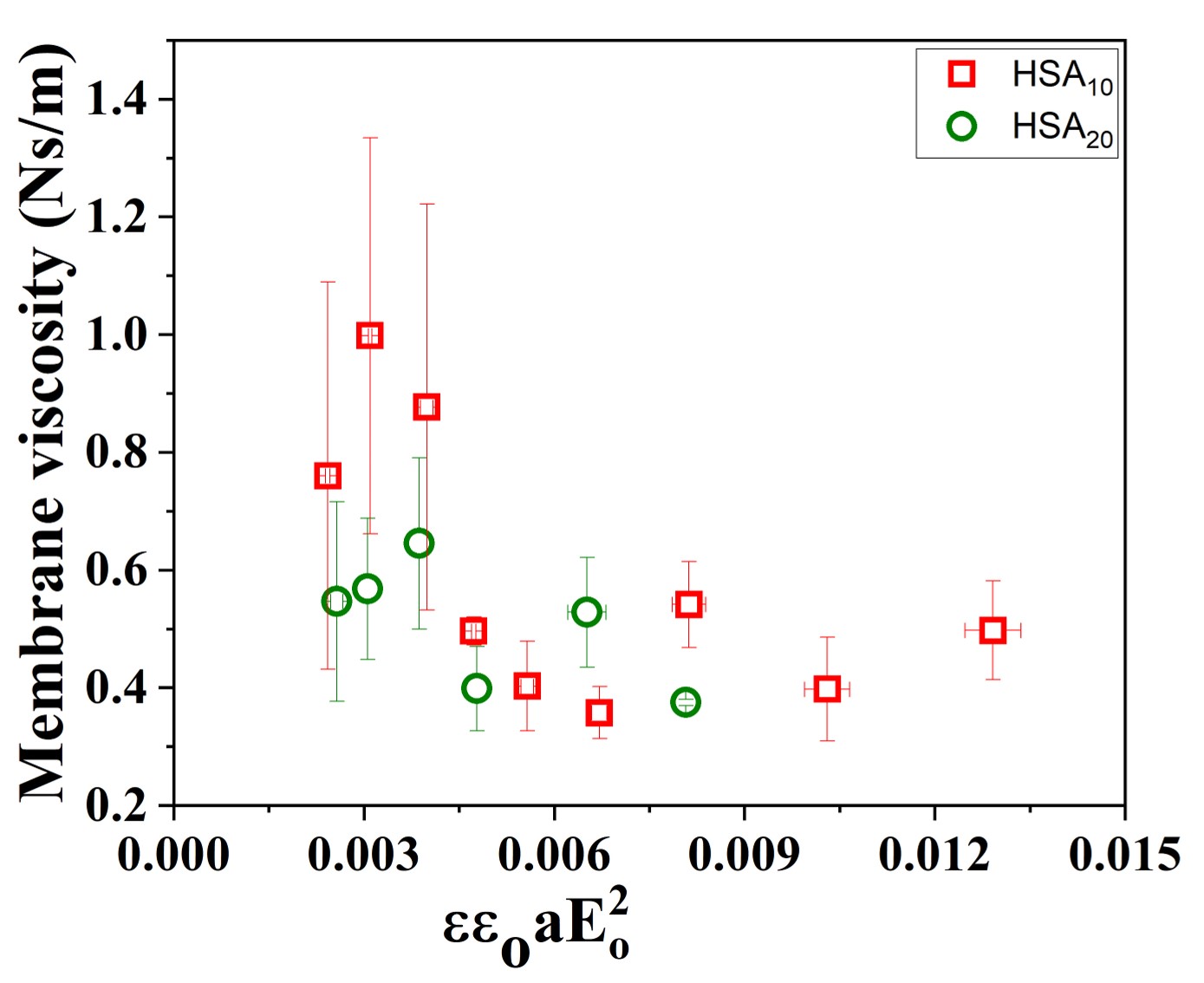}
 		\caption{}
 	\end{subfigure}
 	\begin{subfigure}{0.5\linewidth}
 		\centering
 		\includegraphics[width=1\linewidth]{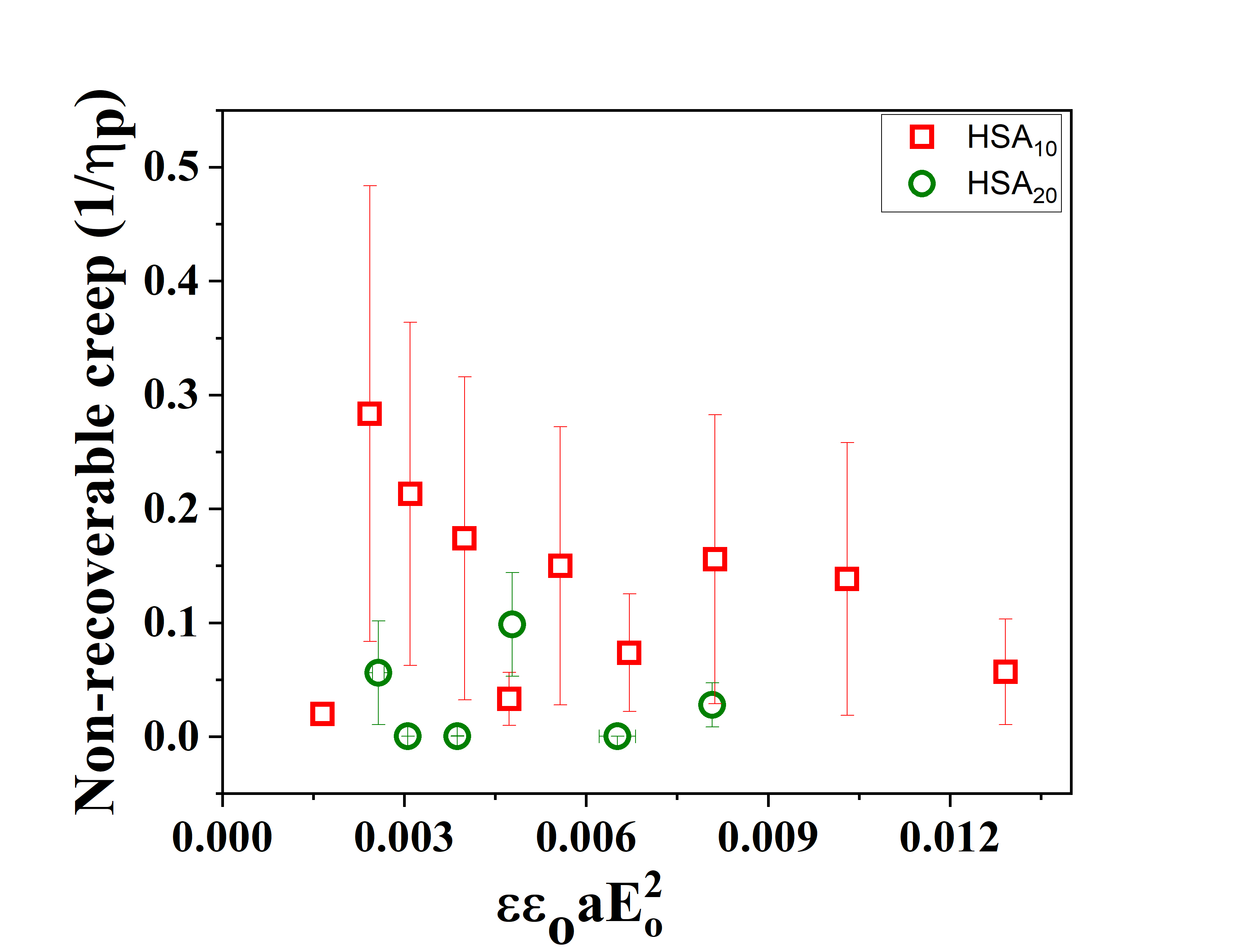}
 		\caption{}
 	\end{subfigure}
 	\caption{ Interfacial rheological properties (a) elasticity, (b) membrane viscosity, and (c) unrecoverable creep at different stresses obtained from model fitting }
 	\label{fig:result_creep_para_with_stress}
   \end{figure}

 \begin{figure}
    \centering
     \begin{subfigure}{0.45\linewidth}
          \includegraphics[width=1\linewidth]{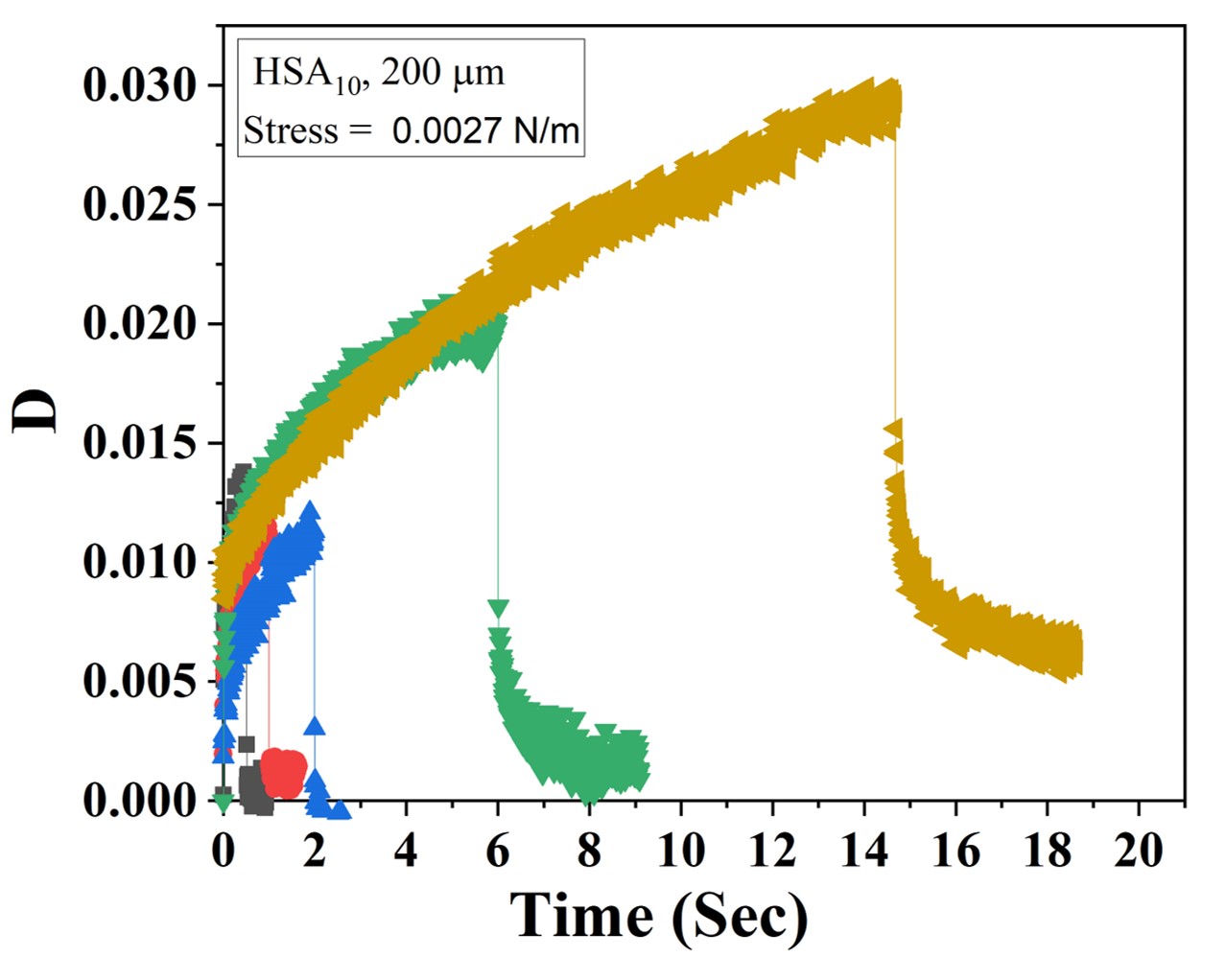}
     \caption{}
     \label{fig:creep_check1}
     \end{subfigure}
     ~
     \begin{subfigure}{0.45\linewidth}
          \includegraphics[width=1\linewidth]{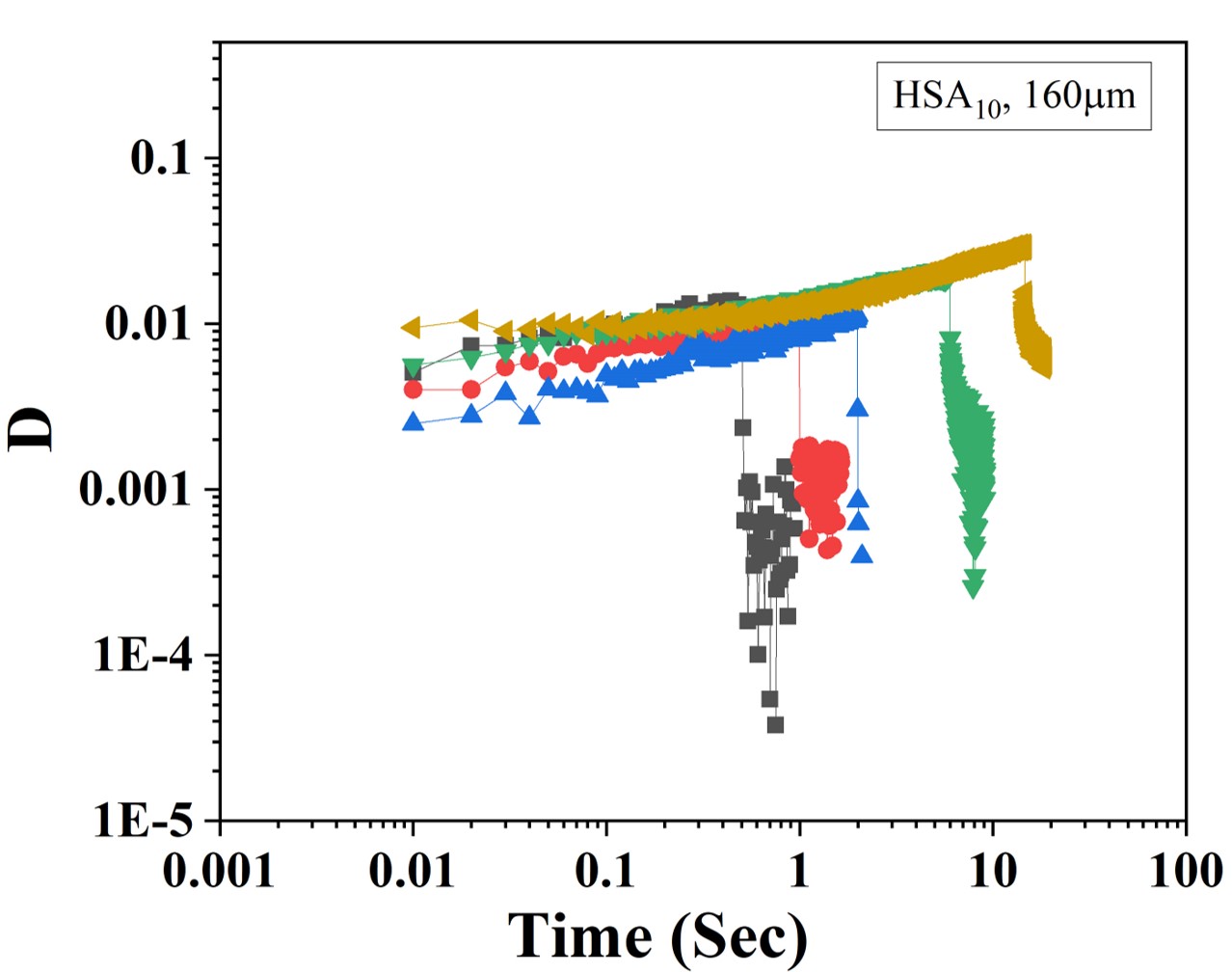}
     \caption{}
     \label{fig:creep_check2}
     \end{subfigure}
     \caption{Dynamic deformation for HSA microcapsules suspended in 350 cSt silicone oil at constant electric field (1kHz) synthesized at pH 7.4 with 10 percent HSA concentration}
     \label{fig:creep_time_check}
 \end{figure}
 
 \begin{figure}
      \centering
     \includegraphics[width=0.7\linewidth]{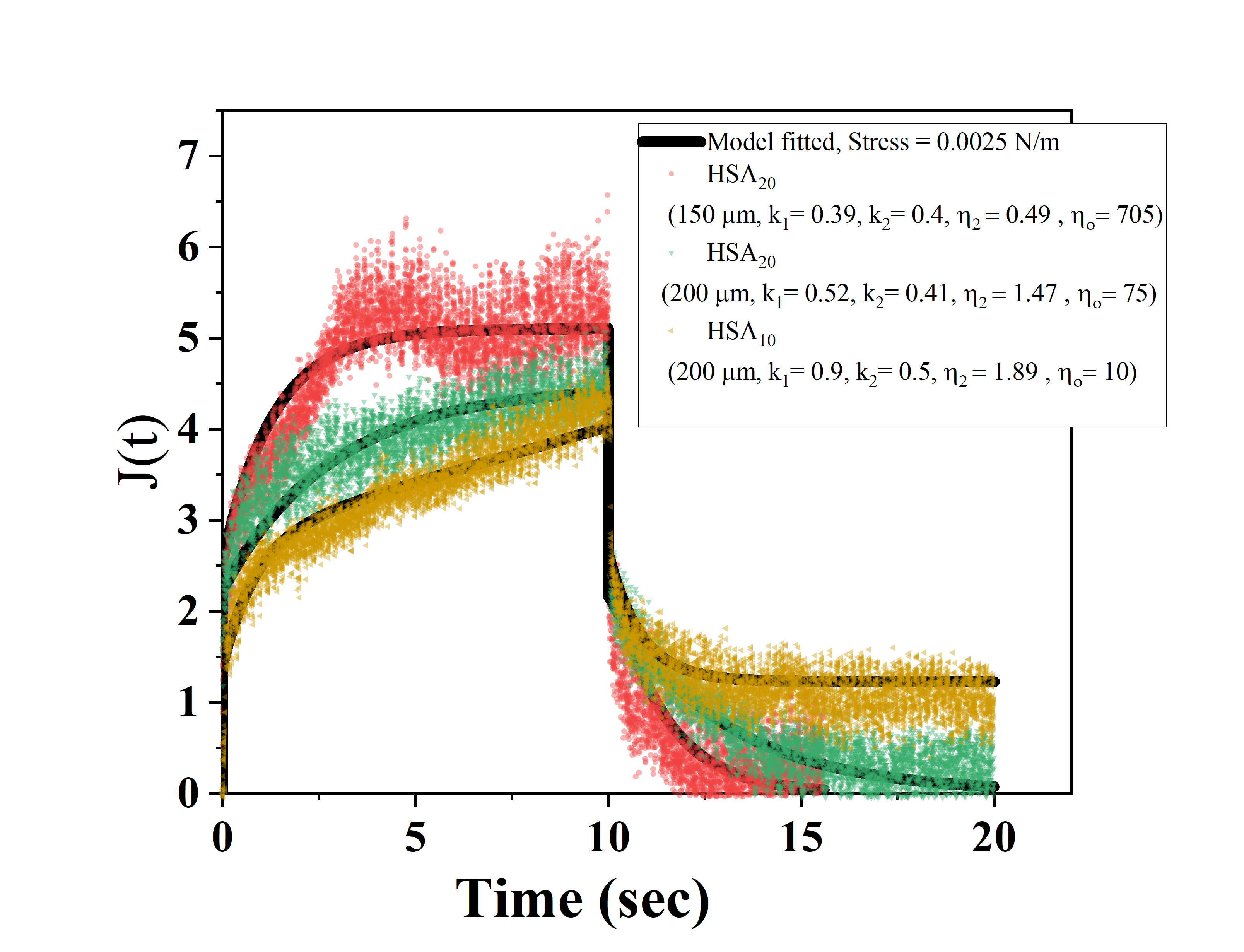}
      \caption{Creep compliance fitted with respect to time for HSA  microcapsules at constant applied electric field at 1 kHz  (solid line shows model fitted) }
     \label{fig:creep_fit}
  \end{figure}

  We first investigated surface elasticity using the approximate analysis method from the small deformation theory. In any typical electrohydrodynamic experiment, a single capsule was suspended in pure silicone oil (350 cSt) in a plastic cuvette, with copper electrodes attached at its inner sidewalls, and a uniform electric field (1 kHz) was applied for about 2 s. As the electric field was switched on, the capsule was deformed to an ellipsoidal shape in the direction of an applied electric field while, after switching off the electric field, it relaxed back as shown in Figure \ref{fig:D_vs_time_HSA}. The dynamics of the deformation was studied at different electric fields until the capsule burst. The temporal data of deformation clearly shows signature of creep (Figure \ref{fig:D_vs_time_HSA}). The surface Young's modulus was calculated using the small deformation theory from the linear regime of deformation versus electric capillary number plot as shown in the Figure \ref{fig:D_vs_Cae}. Since the temporal deformation data shows significant creep, as a preliminary approach, we considered steady deformation value at the end of signal that is at 2 sec for computing surface elasticity by using the Eq. \ref{Ca D0 theory eq_HSA} (See \cite{karyappa2014deformation} for details), which relates the electric capillary number ($Ca_e$) to deformation (D). 
  
Figure \ref{fig:D_vs_Cae} clearly shows that the HSA capsules exhibit strain-softening, for both $HSA_{10}$ and $HSA_{20}$, as indicated by deviation of the $D$ vs $Ca_e$ curve from linearity at higher capillary numbers. This is in agreement with the observation from other studies about HSA capsules  in the literature \cite{gires2014mechanical,gubspun2016characterization}. A more accurate way of analysis is to fit the viscoelastic spring dashpot model to the dynamic deformation curve. Therefore, for each of the experiments in the $D$ vs $Ca_e$ data, a 4 element Burger model was fitted to the temporal data of deformation, and the variation of $E_s,E_y,\eta_m$ and $1/\eta_p$ (unrecoverable creep) are plotted (Figure \ref{fig:result_creep_para_with_stress}). The results indeed show strain softening ($E_s$, $E_y$ decrease with the electric stress).

For a more detailed analysis of the viscoelastic properties of the HSA capsules, we performed creep tests using the electrodeformation method, and the Burger model described previously was fitted to compute their rheological properties. The creep tests were performed at very low electric stress (left side of dotted lines shown in the Figure \ref{fig:D_vs_Cae}) to ensure that the experiments were in the linear viscoelastic regime. To get the estimate of the duration of application of electric stress required to study creep in capsules till it reaches the secondary stage, we applied constant low electric stress for the different duration as shown in Figure  \ref{fig:creep_time_check}. The results indicate that applying stress for 10 sec was sufficient to study creep behaviour. Increasing the time of application of electric field could lead to large deformations, and rendering the system non-linear.

For the creep test, a low electric stress (0.0024-0.0045 $N/m$) was applied to the capsule suspended in 350 cSt silicone oil for 10 s. We continued video recording of electrodeformation experiments for further 10 s to capture creep recovery  after switching off the field. For each experiment, a new capsule was used. The creep compliance J(t), which is the ratio of deformation to stress, is plotted with respect to time for three sample data sets. The five-element spring dashpot model was fitted individually for each capsule using the MATLAB2019a as shown in Figure \ref{fig:creep_fit} and an excellent fit with the theoretical model was observed. The correlation between the fitted model parameters with the viscoelasto-electrohydrodynamic model was used to compute the interfacial rheological properties of HSA capsules. Following this methodology, around 76 capsules were studied using the creep test, and each data point reported was averaged over around 3-8 capsules.

 \begin{figure}[!ht]
 		\centering
 	\begin{subfigure}{0.6\linewidth}
 		\centering
 		\includegraphics[width=1\linewidth]{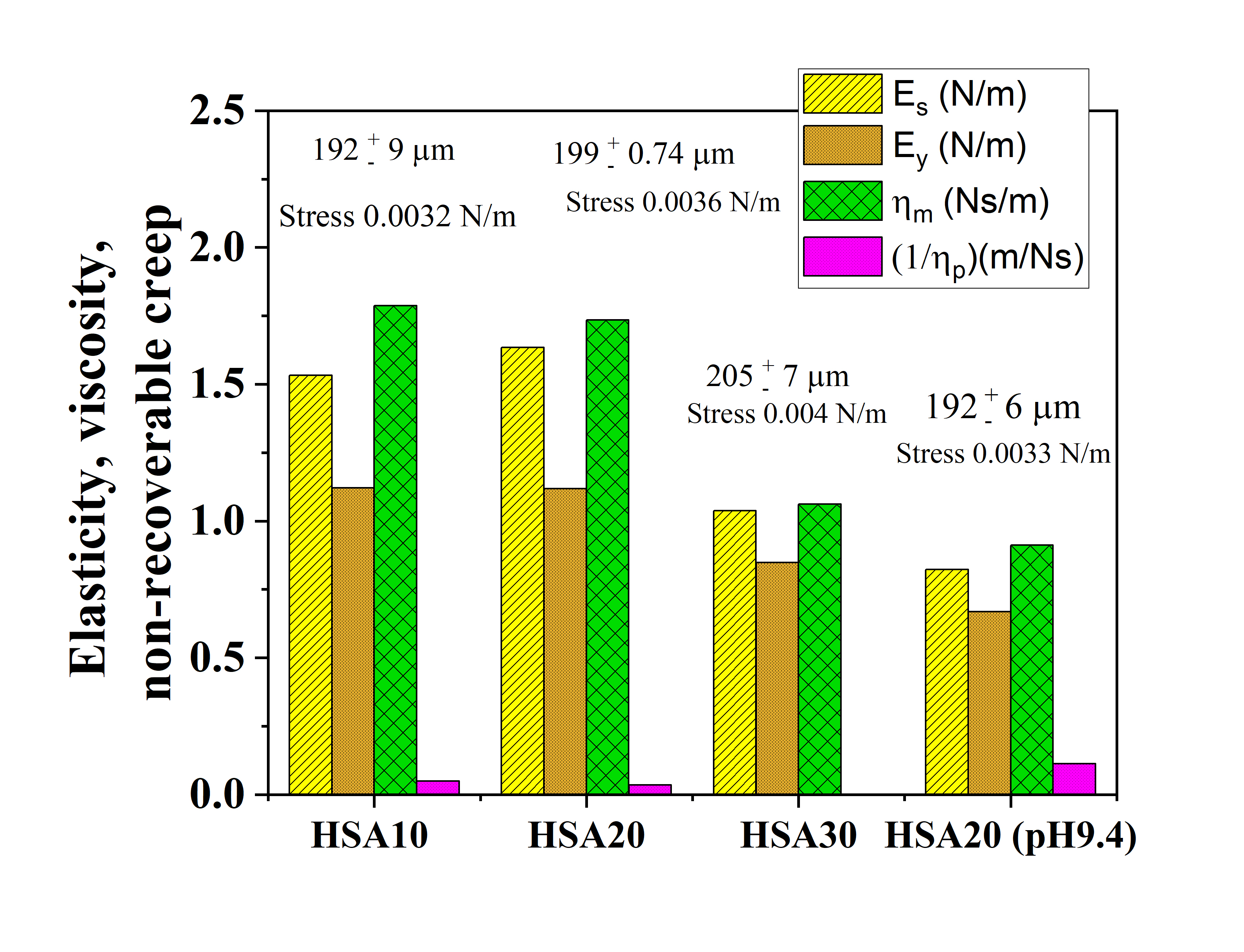}
 		\caption{}
 	\end{subfigure}
 ~	
 	\begin{subfigure}{0.6\linewidth}
 		\centering
 		\includegraphics[width=1\linewidth]{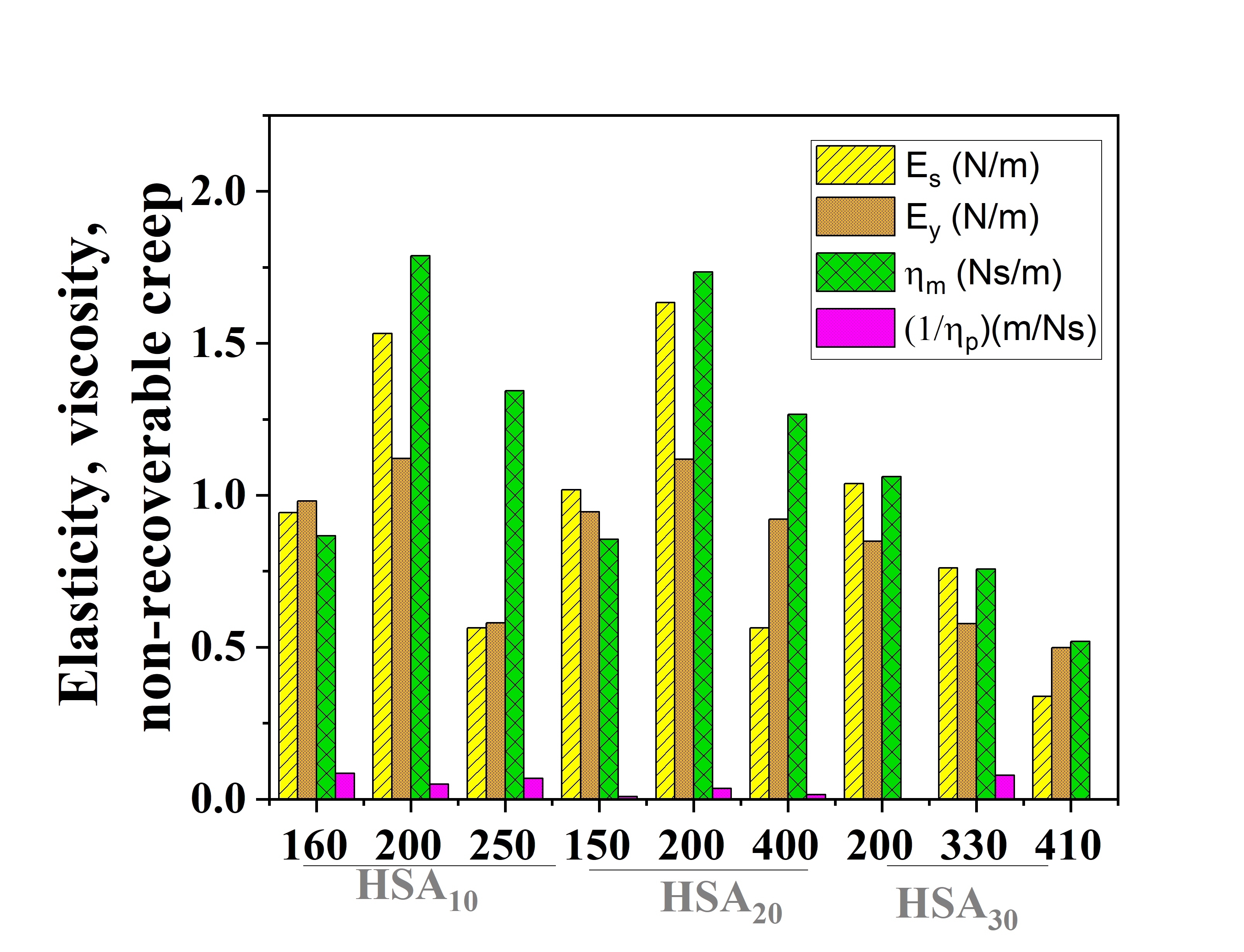}
 		\caption{}
 	\end{subfigure}
 	
 	\caption{ Interfacial rheological properties elasticity ($E_{s}$, $E_{y}$) membrane viscosity ($\eta_{m}$), and unrecoverable creep ($1/\eta_{p}$) obtained from model fitting for (a) capsules synthesized with different HSA concentrations (b) for different capsule sizes and HSA concentrations.}
 	\label{fig:result_creep}
   \end{figure}

 \begin{table}
	\begin{adjustbox}{max width=\columnwidth}
	\begin{tabularx}{\linewidth}{ C *{8}{C} }

\toprule
			\multicolumn{1}{p{3cm}}{\centering HSA concentration \\Radius of capsule ($\mu m$)}
			& $k_1$ 
		      	& {$k_2$} 
			      & {$\eta_2$} & $\eta_o$ \\

			\hline 
			$HSA_{10}$, 160   &  0.335 [0.29-0.43]
 &  0.348 [0.297-0.43] &  0.615 [0.272-1.25] & 8.35 [3.5—]
\\
			\hline 
		$HSA_{10}$, 200   & 0.544
[0.44-0.675] & 0.398 [0.355-0.455]& 1.27
[0.7-2.05]& 14.57 [10.5-25]  \\
			\hline
			$HSA_{10}$, 250	 &  0.2
[0.15-0.34]  &  0.2  [0.175-0.25]  &  0.955 [0.55-1.85] &   10.32 [5.5—]
\\
			\hline 
				$HSA_{20}$ 150 	  &  0.36
[0.26-0.53] & 0.336 [0.265-0.46] & 0.67 [0.13-1.75]&
88.47 [8.5—]
 \\
			\hline 
			$HSA_{20}$ 200 	&   0.544
[0.44-0.68] & 0.3983 [0.355-0.455] & 1.271 [0.7-2.05] &
14.572 [10.5-26]

 \\
		\hline 
				$HSA_{20}$ 400 	&  0.2 [0.145-0.33] &
0.327 [0.27-0.55] & 0.9 [0.57-1.84]& 47.92 [8—]
\\
		\hline 
				$HSA_{30}$ 200 	&  0.369 [0.285-0.52] &
0.3 [0.245-0.385] & 0.75 [0.25-1.74] & 40 [15—]
\\
		\hline 
				$HSA_{30}$ 330 	& 0.27
[0.225-0.375] & 0.2
[0.182-0.225] & 0.537 [0.3-0.99] & 9 [5.5-23]
\\
		\hline 
				$HSA_{30}$ 410 	& 0.12 [0.095-0.21] &
0.177 [0.144-0.29] & 0.36888 [0.175-0.895] & 22892
[8—]
\\	
    	\hline 
				$HSA_{20}$ pH 9.2, 200 	&  0.292
[0.245-0.35] & 0.237 [0.214-4.25] & 0.6482 [0.38-0.98] & 6.32 [4.7-8.7]
\\				
	  \bottomrule
\end{tabularx}
\end{adjustbox}
\caption{The values of estimated model parameters for creep test with the confidence bounds in the parenthesis}	
\label{tab:creep_confidence_bounds}

\end{table}

Figure \ref{fig:result_creep} shows the rheological properties of HSA capsules obtained by fitting the viscoelasto-electrohydrodynamic model. Capsules $HSA_{20}, pH9.2$ synthesized with pH 9.2 buffer show low elasticity and low membrane viscosity than for the capsules $HSA_{20}$ synthesized with phosphate 7.4 buffer with the same HSA concentration.
The FTIR spectra suggests that there is greater reduction in $\alpha$-helicity and a corresponding increase in the $\beta$-sheets in the case of $HSA_{20}$, pH $9.2$. This is possibly due to $HSA_{20}, pH9.2$ being in the basic state, wherein it is highly charged, leading to disruption of the quarternary and tertiary structure. The higlhy charged HSA at pH 9.2, having greater $\beta$-sheets, could lead to increased hydrophobicity, that can cause oligomarisation in the bulk. This could result in the oligomers being weakly crosslinked at the interface, yielding lower values of the rheological parameters. This is supported by the observation of highly porous membrane in the SEM images. 

We next study the effect of concentration on the rheological properties at pH 7.2. 
Figure \ref{fig:result_creep} a shows that the surface elasticity ($E_{s}$), the membrane elasticity ($E_y$), the membrane viscosity ($\eta_m$), and the residual deformation (unrecoverable creep) ($1/\eta_o$) for $HSA_{10}$ and $HSA_{30}$. It is found that although the concentration of HSA proteins is higher for $HSA_{30}$, all the rheological properties are lower than those of $HSA_{10}$. This is counter-intutive, since at higher concentrations, one would have expected higher amount of crosslinking. This is infact hinted by highly folded membranes in $HSA_{30}$ as seen from the SEM images.  However, the SEM images of $HSA_{30}$ also show signficant porosity. The microscopic picture that emerges therefore suggests that the $HSA_{30}$ in bulk water undergoes oligomerisation, leading to larger aggregates at the interface, with high porosity, resulting in capsules with much lower rheological properties. 

The rheological properties of $HSA_{20}$ are similar to values of that of $HSA_{10}$. However, the morphology of $HSA_{20}$ is clearly more rough and folded, and even exibits few pores. This suggests that the rheological properties of $HSA_{10}$ and $HSA_{20}$ could be similar because of two conflicting effects in $HSA_{20}$. While the crosslinking could increase because of higher bulk and thereby surface concentration of $HSA_{20}$, the partial bulk oligomerisation of $HSA_{20}$, a process akin to that discussed for $HSA_{30}$, could result in weak self assembly and weak crosslinking of these oligomers at the interface, resulting in a weakly elastic porous membrane. These two conflicting effects could result in $HSA_{20}$ having similar rheological properties as $HSA_{10}$.

Figure \ref{fig:result_creep}b shows rheological properties of $HSA_{10}$, $HSA_{20}$ and $HSA_{30}$ capsules for different sizes. Table \ref{tab:creep_confidence_bounds} shows the values estimated model parameters and the confidence bounds for all the fitted parameters for creep test.

For both $HSA_{10}$ and $HSA_{20}$, it is observed that the rheological properties, increase from a size of around 150 $\mu m$ to 200 $\mu m$, whereafter, they show a decrease. This could be attributed to an increase in interfacial concentration with increase in size of the capsule. As the size of the capsule increases, the interfacial HSA concentration (molecules per unit surface area) also increases. An increase in surface concentration results in higher interfacial concentration and corresponding crosslinking of the interface, resulting in higher values of rheological properties. Table \ref{tab:surface_conc} shows the total interfacial concentration for 10,20, and 30 $\%$ HSA, for capsule sizes of $150, 200, 250,300,350$, and $400$ $\mu m$. At still higher interfacial concentrations of HSA, oligomeric particles as well as a multilayered film can be formed. Moreover, the unfolding of proteins can be hindered at higher interfacial concentrations due to saturation of the interface.  This can lead to reduced rheological properties.

\begin{table}[!ht]
\centering
	\begin{adjustbox}{max width=\columnwidth}

\begin{tabular}{c c c c}
		\hline 
			Radius of capsule ($\mu m$)
			& \multicolumn{1}{p{3cm}}{\centering $HSA_{10}$\\ (100 mg/ml)} 
		      	& \multicolumn{1}{p{3cm}}{\centering $HSA_{20}$\\ (200 mg/ml)}  
			      & \multicolumn{1}{p{3cm}}{\centering $HSA_{30}$ \\(300 mg/ml)}   \\
			\hline 
	\centering
	    & \multicolumn{3}{c}{$\Gamma_s $ (molecules per $ nm^2$)} \\  \cline{1-4} 
			150   &  45.62 &  91.25 &  136.88 \\
			\hline 
		200    & 60.83 & 121.67& 182.51   \\
			\hline
			250 	 & 76  &152 & 228  \\
			\hline 
			300 	  & 91.25 & 182.51  & 273.77  \\
			\hline 
			350 	& 106.46  & 212.93   & 319.4 \\
			\hline 
			400 	& 121.67 & 243.35  & 365 \\
			\hline 
	\end{tabular}
	  
\end{adjustbox}
\caption{The values of interfacial HSA concentration  available ($\Gamma_s $) (molecules/$nm^{2}$) for droplets of different sizes and bulk HSA concentrations}	
\label{tab:surface_conc}

\end{table}

\subsection{Compararison of rheological properties of HSA capsules from the literature studies}

Table \ref{tab:HSA_lit} shows a comparison of the viscoelastic properties of albumin capsules computed by different techniques as described in the literature.
Table \ref{tab:HSA_lit} shows that the elasticity increases with HSA concentration and the size of HSA capsules, which is opposite to the trend observed in our study. The difference in the results from other studies might be because studies described in the literature were done on the smaller size range, water-in-water HSA capsules with glycerol as an outer phase. After synthesizing capsules with the conventional method, they were washed and suspended in glycerol for 24 h which might have caused the morphological changes in the membrane as it is known that even after crosslinking, the HSA capsules can change their morphology when exposed to a different environment. \citet{levy1991fourier} showed the effect of soaking highly crosslinked microcapsules obtained at pH 9.8  in a slightly alkaline buffer and hydroxylamine to destroy ester bonds, which increased the size of capsules and reduced roughness. Also, the capsules studied for the water-in-water system could be softer with low membrane viscosity. 


 With a higher protein concentration, although the surface concentration is high, the interfacial adsorption and unfolding phenomena are affected by the presence of the neighboring protein molecules. At a very high protein concentration, the adsorption of protein at the interface will be faster, while in comparison the unfolding might be slower. Unfolding of protein at the interface can occur if the kinetics of adsorption is similar or slower than the kinetics of unfolding \citep{wierenga2006adsorption,yano2012kinetics}. 
 Thus this could explain why while at lower concentrations, the elasticity could be lower and capsules smoother due to low bulk concentration of proteins, at very higher bulk concentration of proteins, the surface concentration could be so high that they do not unfold, and thereby the elasticity is again low. This may result in a maxima at some intermediate bulk concentration of proteins. Similarly the elasticity is expected to be higher at higher pH due to participation of carboxylic and hydroxyl groups in reaction with TC.

\begin{table}[!ht]
\begin{adjustbox}{max width=\columnwidth}
\Huge
\begin{tabular}{ c c c c c c c}
		\hline 
		
		\multicolumn{1}{p{5cm}}{\centering capsule diameter \\ ($\mu m$)} &  \multicolumn{1}{p{5cm}}{\centering HSA concentration\\($\%$) }  & \multicolumn{1}{p{5cm}}{\centering Elasticity  (N/m)} & \multicolumn{1}{p{5cm}}{\centering membrane viscosity (Ns/m)} & Technique & Model used &  Reference \\
			\hline 
			 140-310  &   $5-10-15\%$ & 0.002-5  & - &   Elongational flow & \multicolumn{1}{p{5cm}}{\centering small deformation theory }   &  \cite{de2014mechanical} \\
			\hline 
	 30-80  &  15-25 & 0.007-0.016  & - &  Elongational flow & \multicolumn{1}{p{5cm}}{\centering Generalized Hookes law\\ (Poisson ratio 0.4) }  &  \cite{de2015stretching}\\
			\hline
				 40-450  &   15-35 & 0.02-1  & - &   Elongational flow & Neo-Hookean  &  \cite{gubspun2016characterization} \\
			\hline 
		 110-124 &   20-30 & 0.02-0.12  & 0.0008 & shear flow & Neo-Hookean  &  \cite{gires2016transient}\\
			\hline 
	 54 & 20-25 &  - & 0.00013-0.00025 &\multicolumn{1}{p{5cm}}{\centering tank treading\\ shear flow }  & -  &  \cite{de2016tank}\\
			\hline 
		 110-124  &   20 & 0.002-0.02 & 0.0008 & shear  & Neo-Hookean  &  \cite{gires2014mechanical}\\
			\hline 

	\end{tabular}
\end{adjustbox}
\caption{comparison of the viscoelastic properties of albumin capsules computed by different techniques\\
(Note: All studies are reported in this table with glycerol as outer fluid, with 30 min reaction time at pH 8 except the study by \citet{gires2014mechanical}, which is with dragozat as outer media and 60 sec reaction time at pH 9.8) }	\
\label{tab:HSA_lit}
\end{table}

 \subsection{Frequency sweep test results:}
 
  \begin{figure}[!ht]
     \centering
     \includegraphics[width=0.9\linewidth]{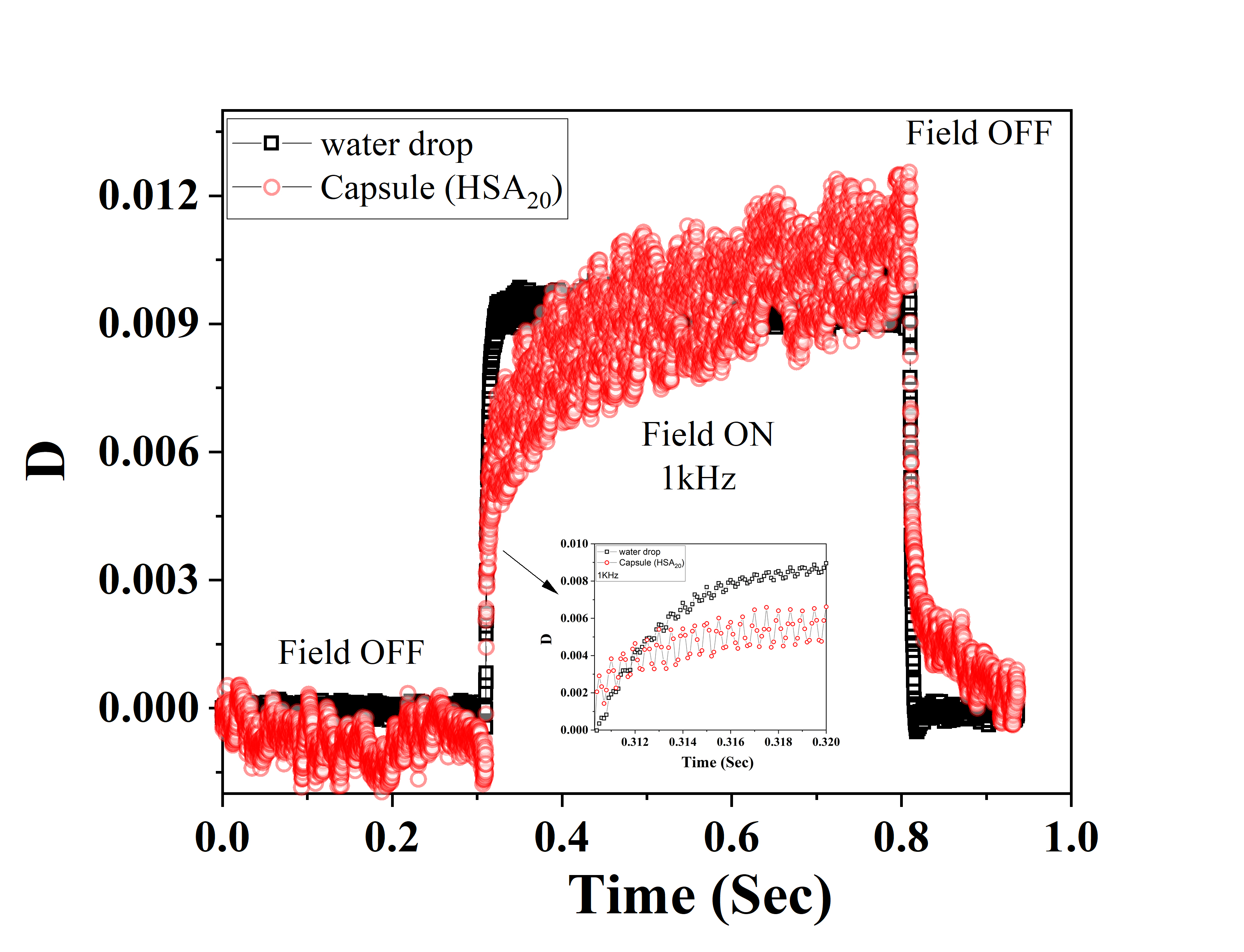}
     \caption{Dynamic deformation of drop and capsules place in the same cuvette containing 350 cSt silicone oil under constant electric field at 1kHz frequency.}
     \label{fig:Drop_capsules_D_vs_t_1KHZ}
 \end{figure}

  \begin{figure}[!ht]
 		\centering
 	\begin{subfigure}{0.45\linewidth}
 	\includegraphics[width=1\linewidth]{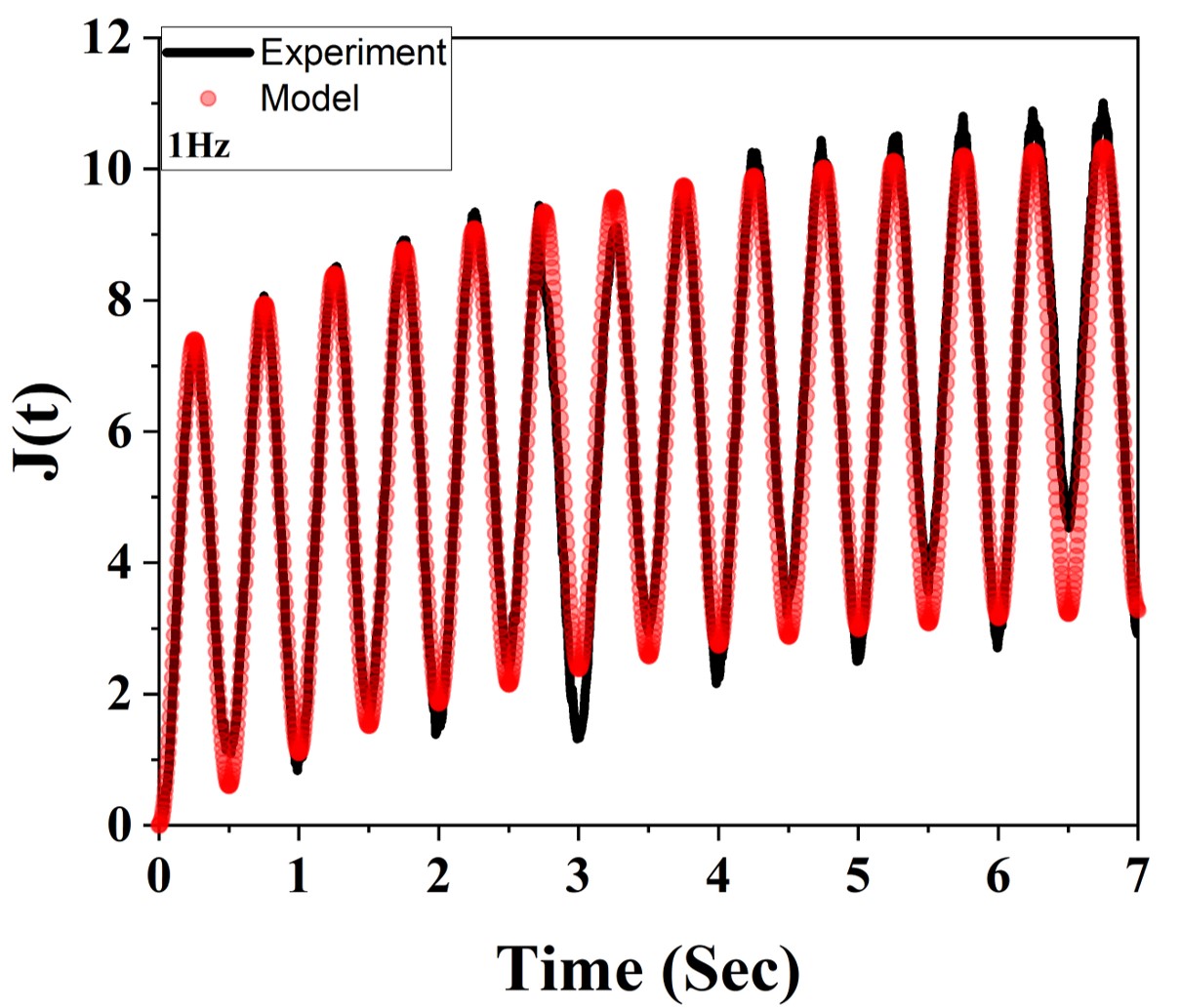}
 	\caption{}
 	\label{fig:osccilatoryfit}
 	\end{subfigure}
 	~
 	\begin{subfigure}{0.45\linewidth}
 	\includegraphics[width=1\linewidth]{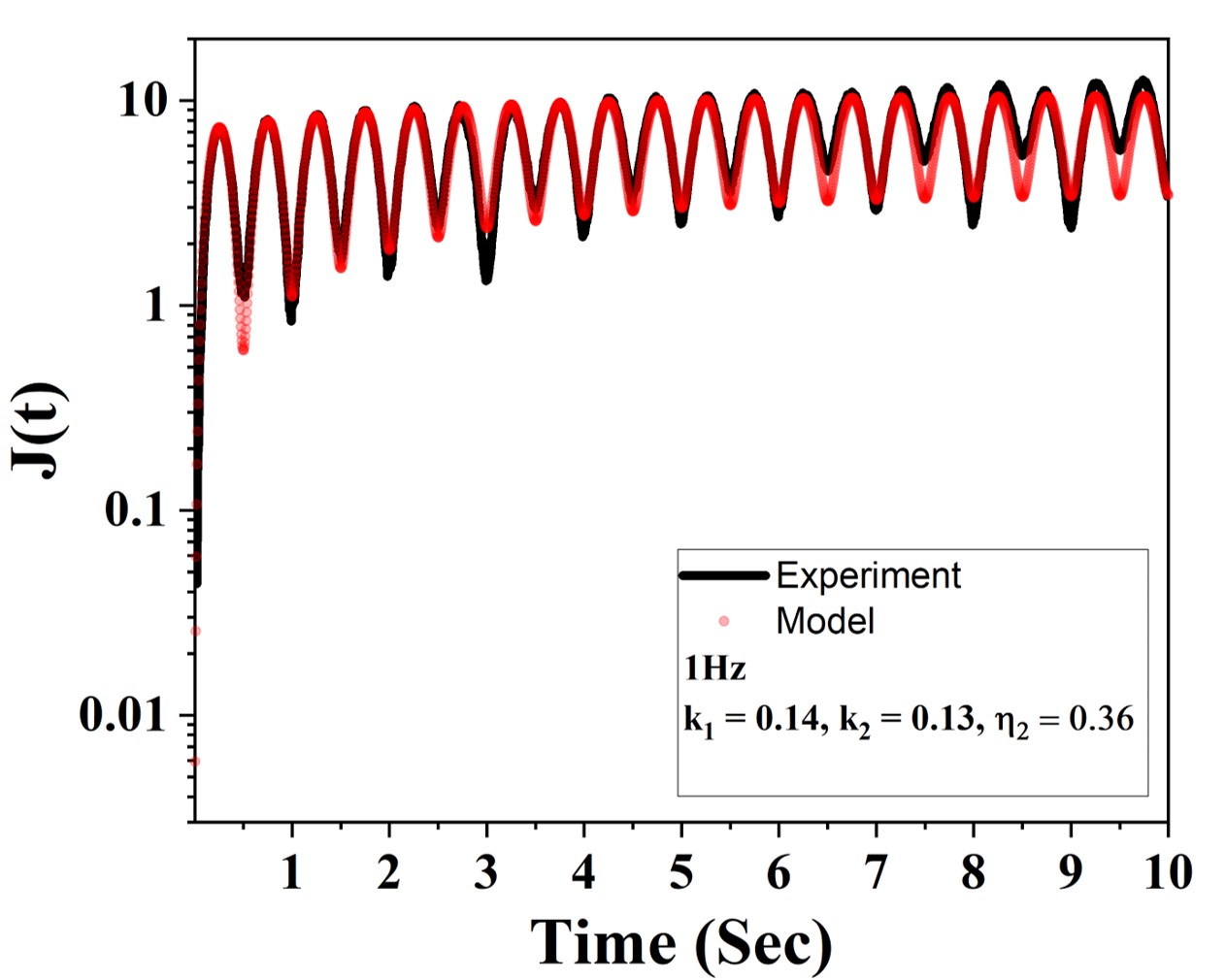}
 	\caption{}
 	\label{fig:osccilatoryfitlogscale}
 	\end{subfigure}
 	\caption{Oscillatory viscoelastic model fitted for (a) $HSA_{20}$ capsule at 10 Hz (b) shows a same plot with semilog scale}
 	\label{fig:frequency_sweep_results}
 \end{figure}

The oscillatory test was conducted on the $HSA_{20}$ capsule suspended in 350 cSt silicone oil using the electrodeformation method described earlier. The smaller water drop was also suspended in the same cuvette for reference to get the exact estimate of time t=0, at which the field was switched on. Figure \ref{fig:Drop_capsules_D_vs_t_1KHZ} shows dynamic deformation of drop and capsules placed in the same cuvette at 1 kHz. The deformation of water drop is scaled to match with capsule deformation in Figure \ref{fig:Drop_capsules_D_vs_t_1KHZ}. As the sinusoidal electric field was applied, the capsules showed oscillatory deformation. 
 A constant low electric stress (0.0025 N/m) was applied at different frequencies varying from 10 Hz to 1 kHz, and the deformation video was recorded to study the effect of frequency on membrane viscoelastic properties. The dynamic deformation was fitted (Figure \ref{fig:frequency_sweep_results}) with the model described earlier, and the membrane surface elasticity and viscosity were computed from the model parameters.

 \begin{figure}[!ht]
    \centering
    \includegraphics[width=0.9\linewidth]{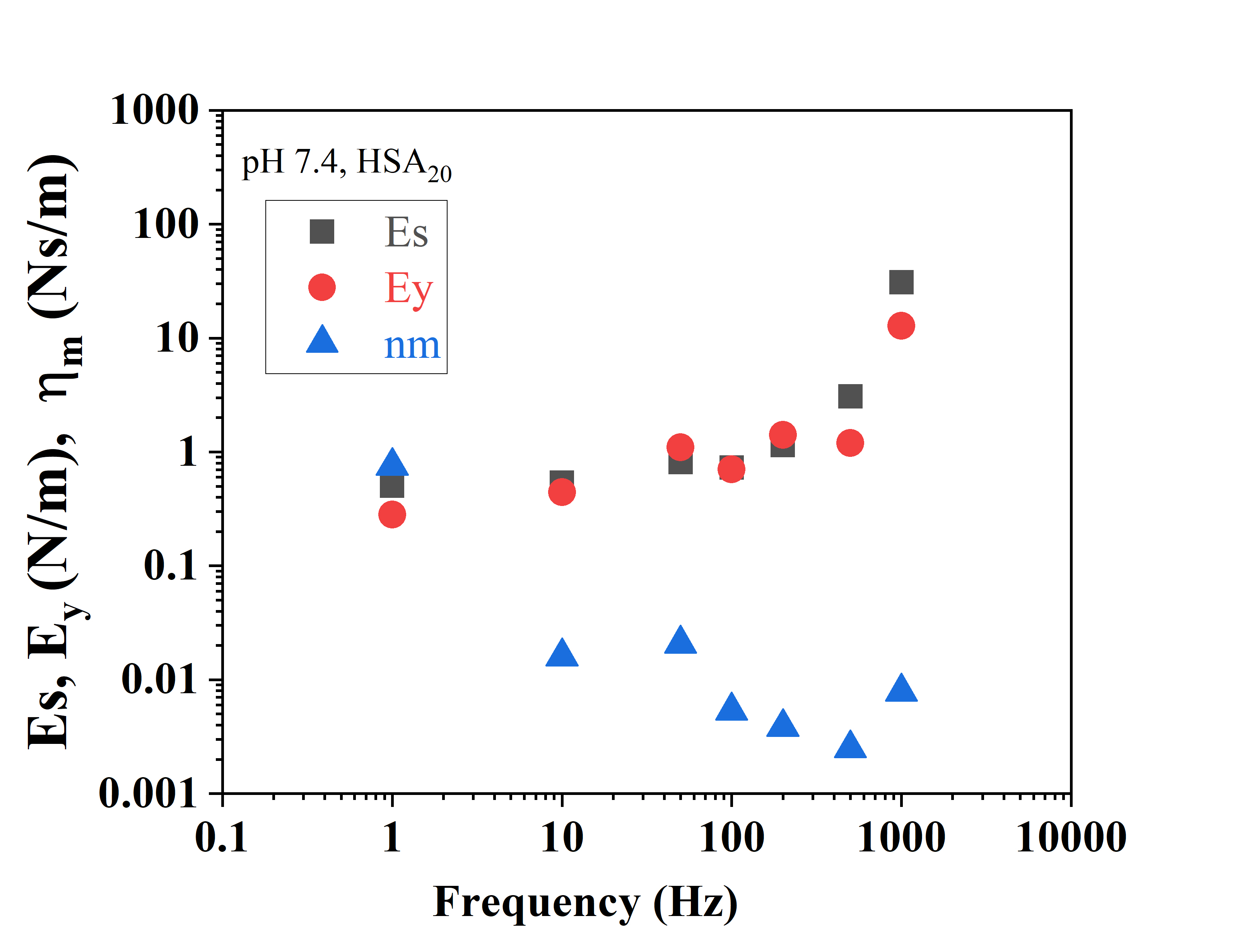}
    \caption{Variation of elasticity and membrane viscosity of $HSA_{20}$  capsules  with frequency}
    \label{fig:frequency_results}
\end{figure}

 Figure \ref{fig:frequency_results} shows that as the frequency increases, the elastic modulus increases, indicating the dominance of elastic response. There are two inherent time scales in the interfacial rheology of HSA capsules, $\tau_1$ and $\tau_2$, which are of the order of $1s$ and $1 ms$, thereby suggesting a crossover in properties around 1 Hz and 1 kHz. 
 
 With an increase in frequency, the timescale for viscous relaxation processes decrease, and thus the elastic response (storage modulus) shows dominance over the loss modulus. At 1 Hz, which is the lowest frequency used in the present study, both elasticity and viscosity are approximate of the same order indicating crossover frequency. This suggests that there could be irreversible structural changes in the interfacial film at lower frequencies resulting in structural breakup,  which suggests at using still smaller applied stress. 
 
 \begin{table}[!ht]
	\begin{adjustbox}{max width=\columnwidth}
	\begin{tabularx}{\linewidth}{ c *{4}{C} }

\toprule
			Frequency ($Hz$)
			& $K_1$ 
		      	& $K_1$ 
			      & $\eta_2$ & $\eta_o $ \\
	  
			\hline 
			1  & 0.18 [0.144-0.239]
  &  0.1 [– -1.3] &
0.5363
[0.265-2.3]&
29612
[0.9—]
 \\
			\hline 
		10    & 0.19
[0.147-0.275] &
0.158
[0.111-0.252] &
0.011
[0.00289-0.0389]&
12637
[0.146–]
  \\
			\hline
			50 	 &  0.288
[0.22-0.418]&
0.389
[0.105-1.9]&
0.0147
[0.0014-0.11]& 5585.044
[0.038–]

   \\
			\hline 
			100 &  0.258
[0.228-.3]&
0.249
[0.186-0.345]&
0.0038
[0.0019-0.0066]&
61104
[-0.038- -]
\\
			\hline 
			200 	&  0.4086
[0.362-0.467]&
0.5
[0.383-0.685]&
0.0027
[0.0011-0.0054]&
82643
[0.035- -]
\\
			\hline 
			500 	&   1.093
[0.675-2.66]&
0.4249
[0.245-0.78]&
0.00172
[0.00045-0.0043]&
41513
[0.012- -]
\\
			\hline 
			1000 	&  10.93
[7.48-20.5]&
4.54
[3,37-3.55]&
0.0055
[0.0011-0.0148]&
157015
[0.06 - – ]
\\
	  \bottomrule
\end{tabularx}
\end{adjustbox}
\caption{The values of estimated model parameters for oscillatory test with the confidence bounds in the parenthesis}	
\label {tab:phase lag_parameter}
\end{table}

 \section{Summary and conclusion:}

 The configuration of even uncrosslinked protein in the bulk solution is complex and dependent on parameters such as pH and salinity. The behavior of protein at the oil-water interface is even more complex due to the arrangement of the hydrophobic and hydrophilic parts at the oil-water interface. Therefore, it is apparent that the microstructure of crosslinked HSA at the oil-water interface could have a hierarchy of structures. The FTIR studies, although conducted on an enseumble of microcapsules of varying sizes, indicates a significant increase in the $\beta$ sheet content at higher pH. The SEM studies indicate smoother smaller capsules, while capsules at higher pH and higher concentrations are rough. Additionally they may also have pores. EHD based interfacial rheological studies indicate that the creep and oscillatory response of the capsules can be reasonably described by the four-element Burger model, and the capsules can admit viscoelasticity as well as unrecoverable creep.Creep experiments performed on HSA capsules show that at a higher concentration of HSA, membrane with less elasticity is formed, which also shows more unrecoverable creep. Similarly, at higher pH, membranes with poor rheological properties are formed. Likewise, as the size is increased, the interfacial rheology first improves, becoming more viscoelastic, while at larger sizes, the viscoelastic properties reduce. The mechanism at play seems to be higher $\beta$ sheets and thereby higher unfolding and hydrophobic oligomerisation at higher pH, whereby the loose network of oligomers at the interface leads to weaker membranes with lower rheological values at higher pH. The weak oligomeric agglomeration at the interface also leads to a porous membrane. The reduction in viscoelastic properties at higher bulk concentrations for the same size of the capsules, seems to be due to more oligomerisation at higher concentration leading to weakly crosslinked porus membranes at higher bulk concentrations, as well as due to multilayer formation resulting in weaker membranes. The weaker rheological properties, at very larger surface concentrations (larger sizes too) is possibly due to multilayer formation and thereby weakly crosslinked membranes. On the other hand, in the low interfacial concentration regime, as the interfacial concentration increases, the rheological properties improve due to increase interfacial concentration and crosslinking. Thus, the properties seem to be dependent upon the extent of oligomerisation in the bulk; which could depend upon the pH (through hydrophobic interactions and destruction of native structure), interfacial concentration, multilayer formation, and unfolding characteristics at the interface, which depend upon the surface concentration, hydrophobic interactions, crosslinking at the interface, and the dynamics of unofolding vursus adsporption. An oscillatory test was performed in the high-frequency range that shows a dominance of the elastic network at higher frequency, thereby indicating that the system might be yielding at lower frequencies.

   To conclude, this work tries to establish a microstructure-rheology relationship for HSA microcapsules. Understandably, the origin of viscoelasticity in this system, although expected to be non-trivial seems to be represented by the Burger model in the kHz frequency range. The EHD method developed here for explaining high-frequency interfacial rheology therefore shows promise. A more detailed investigation would thus unravel several mysteries in this interesting system, and the work presented here could be considered to be a humble beginning in this regard.

\bibliographystyle{plainnat}
\bibliography{references.bib}

\newpage

\pagenumbering{arabic}
\clearpage

\renewcommand{\thefigure}{S\arabic{figure}}
\renewcommand{\thepage}{S\arabic{page}} 
\setcounter{figure}{0}
\newcommand{\beginsupplement}{%
        \setcounter{table}{0}
        \renewcommand{\thetable}{S\arabic{table}}%
        \setcounter{figure}{0}
        \renewcommand{\thefigure}{S\arabic{figure}}%
     }


\section*{Supplementary Information: Study of Interfacial Rheology of Human Serum Albumin Microcapsules using Electrodeformation Technique}

\subsection*{Capsules size distribution}

 \begin{figure}[!ht]
 	\centering
 	\includegraphics[width=0.7\linewidth]{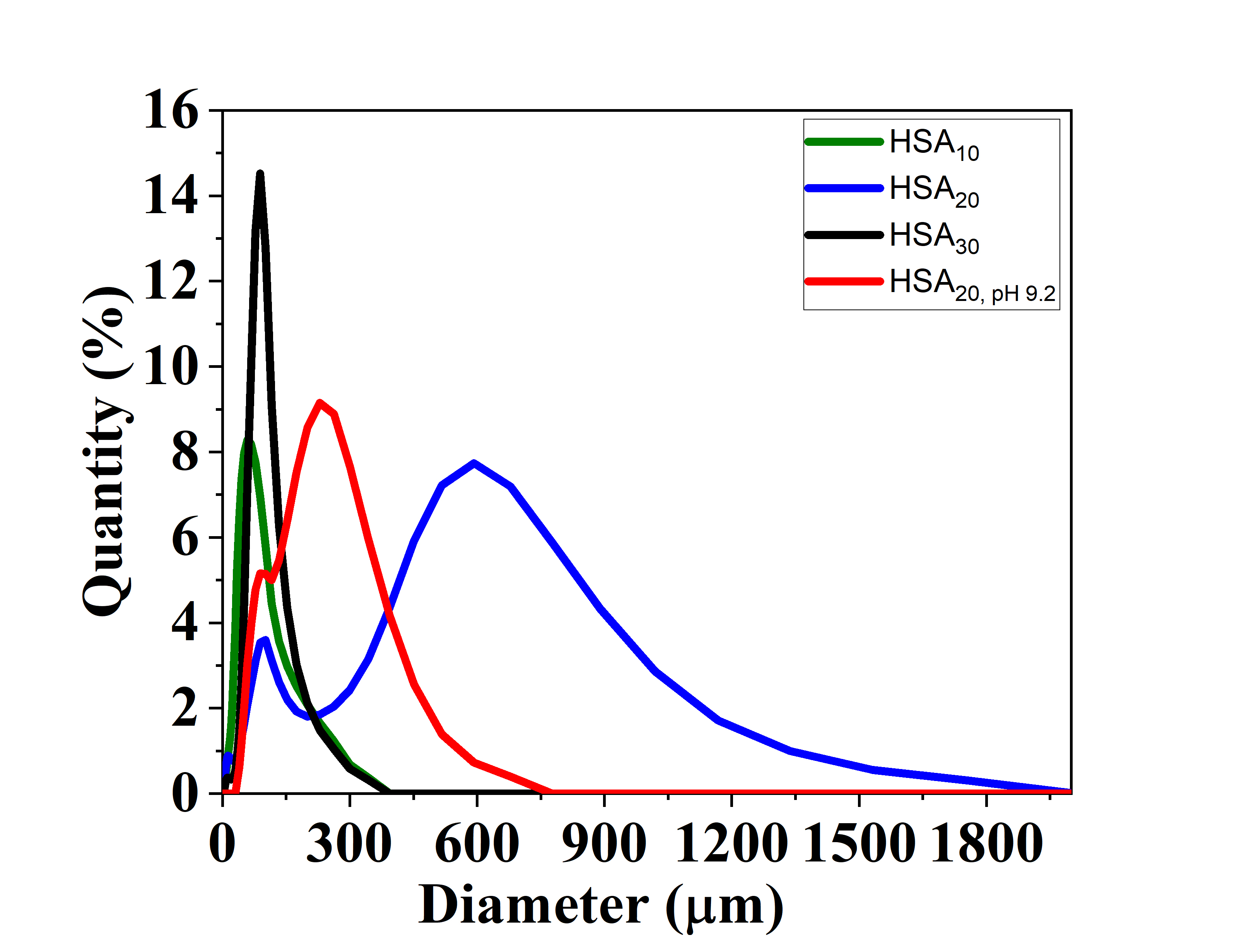}
 	\caption{Size distribution for HSA capsules synthesized at 600 rpm with different HSA concentrations and pH}
 	\label{fig:size_distribution}
 \end{figure}

Figure \ref{fig:size_distribution} shows the size distribution of HSA capsules studied by using a particle size analyzer (Horiba). The outer organic phase (cyclohexane) was replaced by deionized water by decanting the oil phase and successively washing capsules with deionized water. Finally, capsules were suspended in the deionized water for studying particle size distribution. 

While the size distribution for $HSA_{10}$ and $HSA_{30}$ is unimodal, the one for $HSA_{20}$ for both phosphate and carbonate buffers is bimodal. 
The smaller capsules were formed in the range of 100-150$\mu m$ for all the four cases. For $HSA_{20}$, and $HSA_{20,9.2}$, which show bimodal distribution, the $2^{nd}$ peak is seen around a size range of 300 and 600 $\mu m$ respectively. It is unclear if the $2^{nd}$ peak for $HSA_{20}$ is due to agglomeration of capsules or if the disribution is really bimodal. The electrohydrodynamic experiments to determine the rheology of the capsules were conducted on one single capsule at-a-time, chosen from these distributions. It should be remebered that these distributions themselves are for water-in-water HSA capsules, while the EHD experiments were conducted on water-in-oil HSA capsules.  

A number of factors determine the size distribution of these capsules. The size distribution of a water-in-oil emulsion depends upon the interfacial tension and surface elasticity of the resulting droplets. While the system starts as an emulsion of water droplets in oil, coated with HSA and stabilized by surfactant (span 85), it soon undergoes interfacial condensation, which can possibly both prevent breakup as well as prevent coalescence of the resulting capsules. This complex interplay of uncrosslinked droplets and crosslinked capsules, could be responsible for the resulting size distribution of the capsules. While higher concentration of HSA can reduce the intefacial tension, it can also increase the interfacial elasticity of the interface.

\subsection*{Characterization of capsules using Fourier transform infrared spectroscopy (FTIR) technique:}

\begin{figure}[!ht]
 	\centering
 	\includegraphics[width=0.7\linewidth]{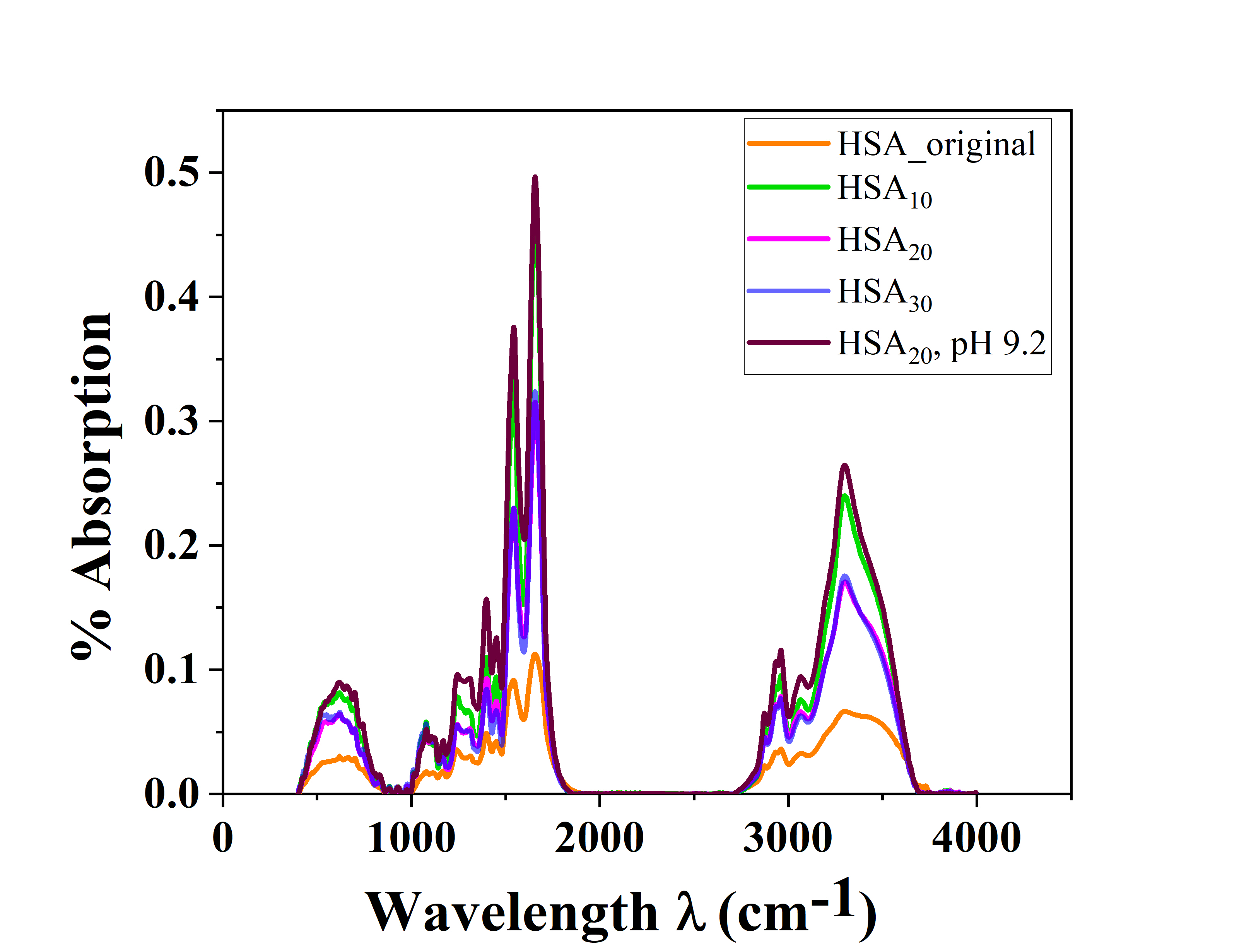}
 	\caption{Full range FIIR spectra for original lyophilized HSA (without crosslinked) and for HSA capsules synthesized at different reaction conditions}
 	\label{fig:ftir}
 \end{figure}
 \begin{figure}[!ht]
 	\centering
 	\includegraphics[width=0.7\linewidth]{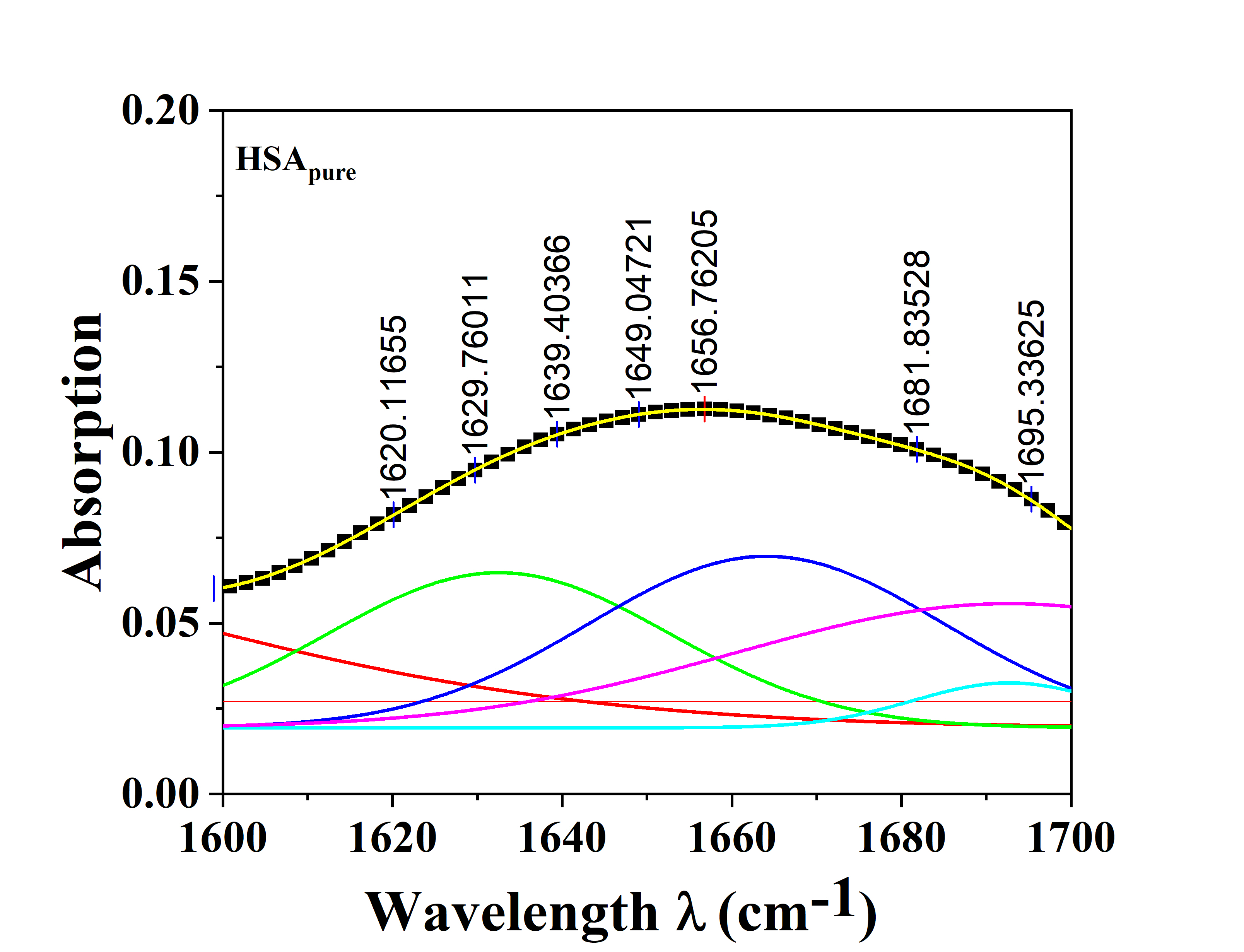}
 	\caption{The Amide I band and its second derivative deconvolution fitted for original uncrosslinked HSA}
 	\label{fig:ftir_o}
 \end{figure}

 \begin{figure}[!ht]
 \centering
 	\begin{subfigure}{0.45\linewidth}
 		\centering
 		\includegraphics[width=1\linewidth]{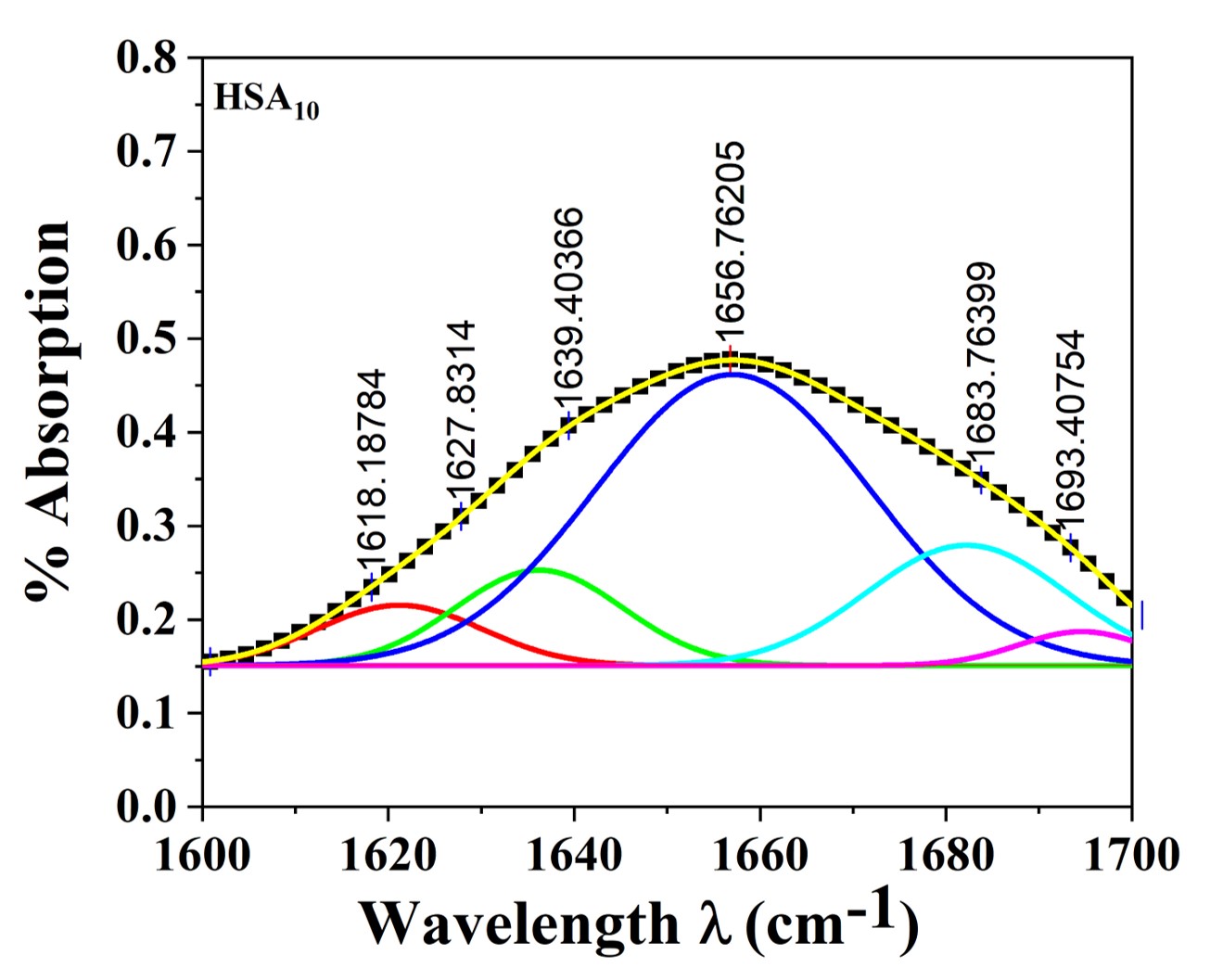}
 		\caption{}
 	\end{subfigure}
 ~	
 	\begin{subfigure}{0.45\linewidth}
 		\centering
 		\includegraphics[width=1\linewidth]{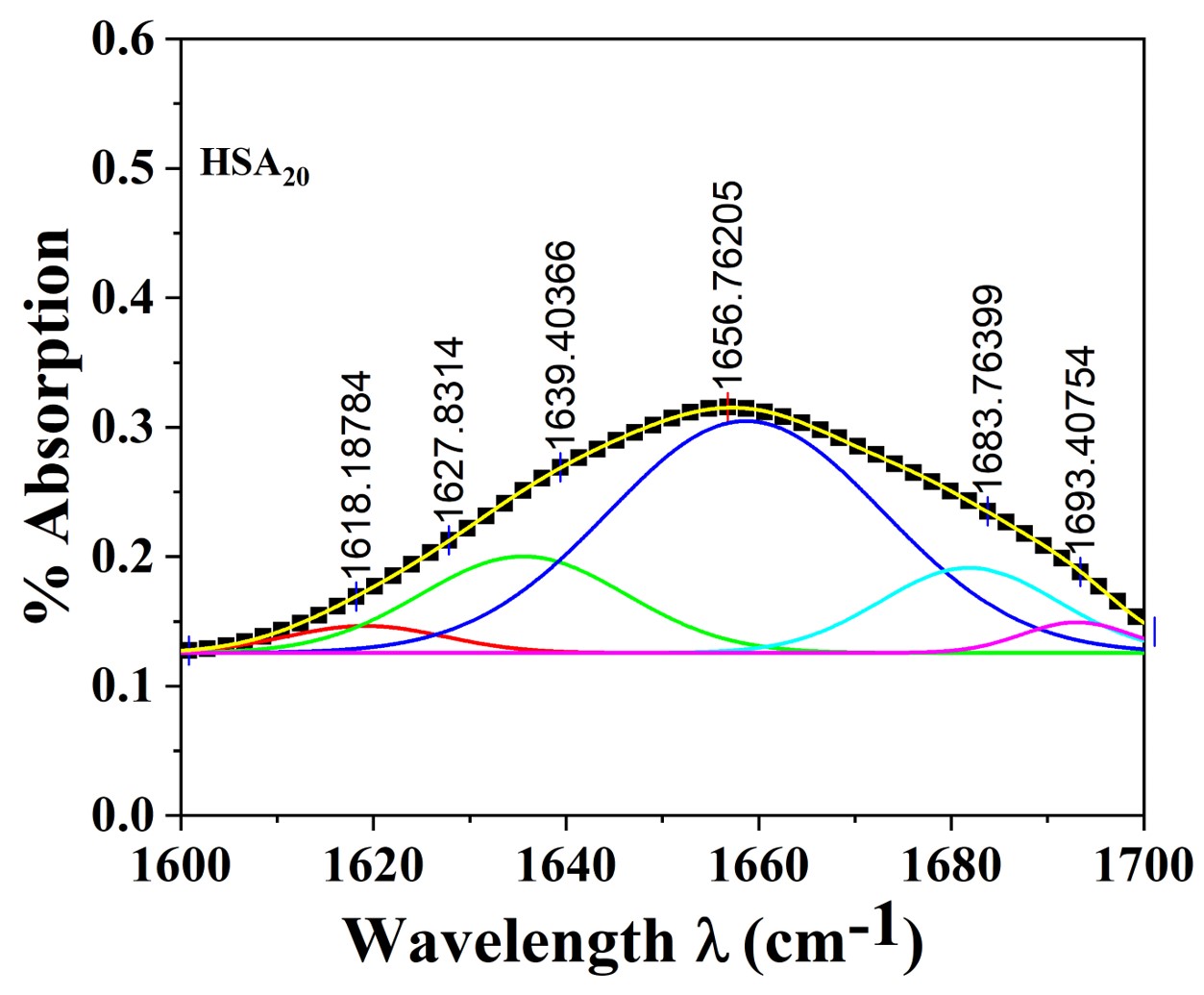}
 		\caption{}
 	\end{subfigure}
 ~	
 	\begin{subfigure}{0.45\linewidth}
 		\centering
 		\includegraphics[width=1\linewidth]{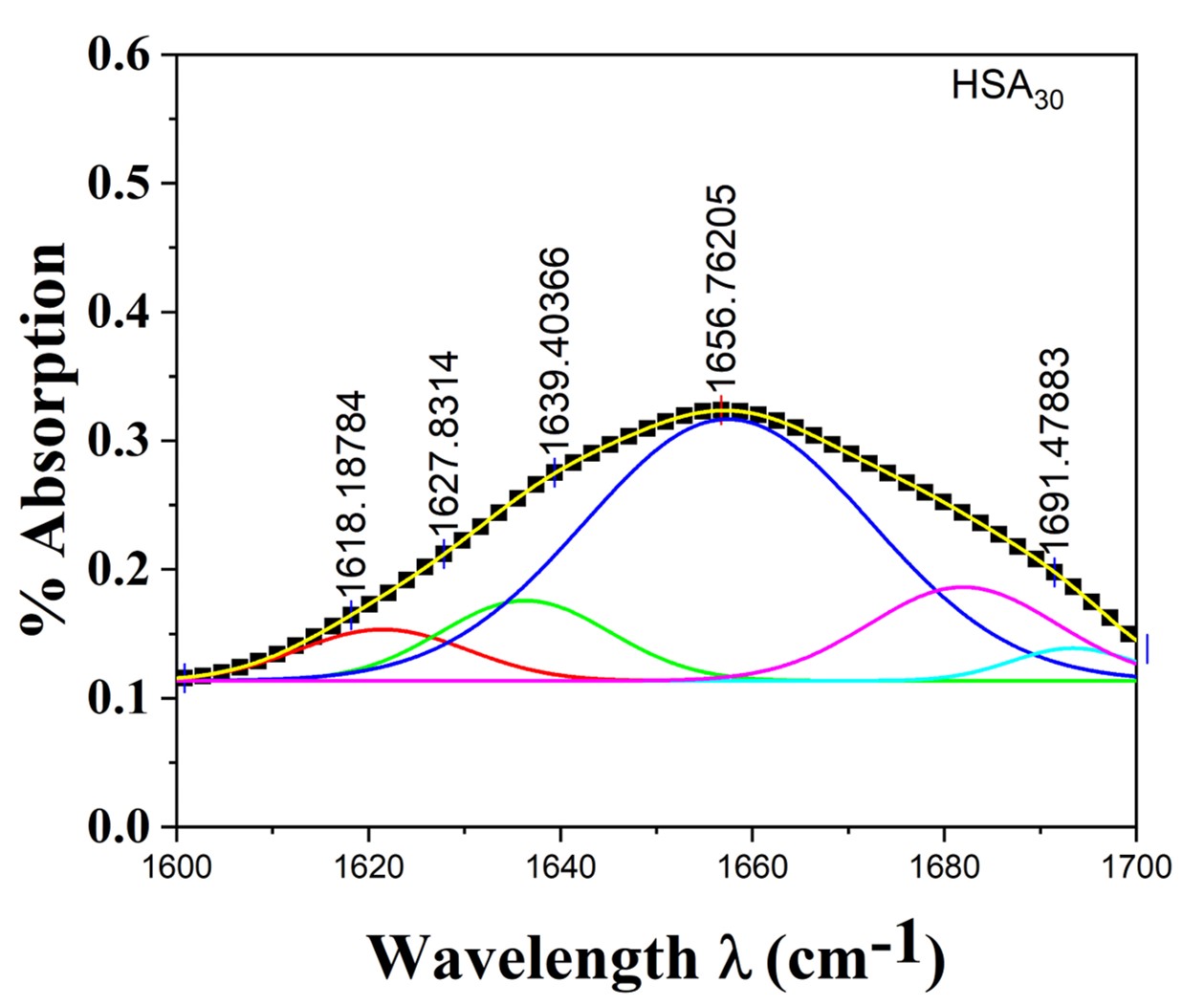}
 		\caption{}
 	\end{subfigure}
 	~
 	\begin{subfigure}{0.45\linewidth}
 		\centering
 		\includegraphics[width=1\linewidth]{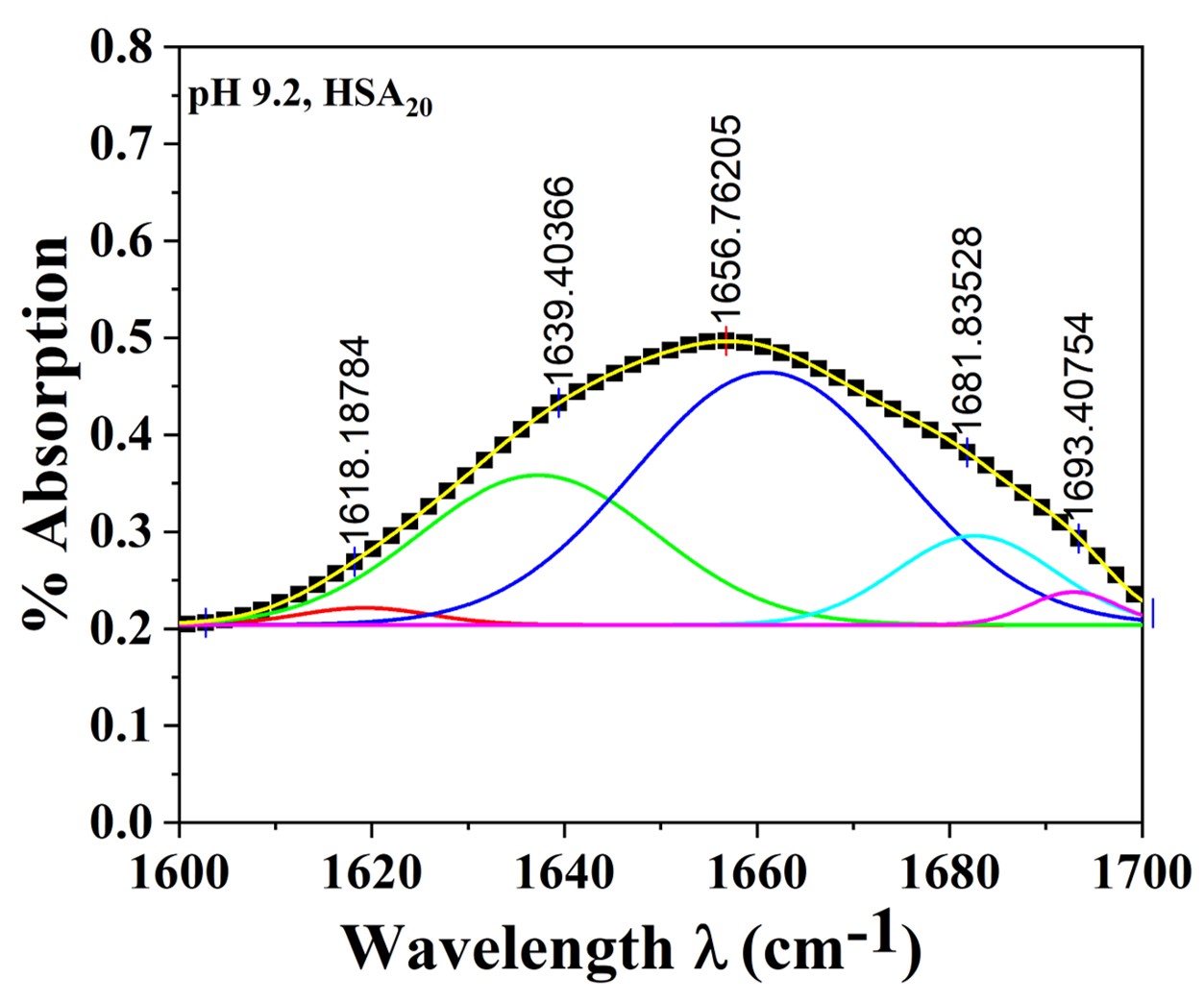}
 		\caption{}
 	\end{subfigure}
 	\caption{The Amide I band and its second derivative deconvolution fitted for capsules prepared with 15 min reaction time with phosphate pH 7.4 buffer (a) $HSA_{10}$(b) $HSA_{20}$ (c) $HSA_{30}$, (d) $HSA_{20},pH 9.2$ }
 	\label{fig:FTIR_fit}
   \end{figure}



\end{document}